\documentclass{aastex6}

\usepackage{amsmath}
\usepackage{graphicx}
\usepackage{hyperref}
\usepackage{ctable}
\usepackage{float}
\usepackage[caption=false]{subfig}
\usepackage{bm}
\usepackage{xcolor}

\newcommand{\bt}[1]{\mathbf #1}
\newcommand{\sk}[1]{}

\newcommand{\refeq}[1]{Eq.~(\ref{eq:#1})}          
          
\newcommand{\reffig}[1]{Fig.~\ref{fig:#1}}          
\newcommand{\refsec}[1]{Sec.~\ref{sec:#1}}
\newcommand{\refapp}[1]{App.~\ref{app:#1}}
\newcommand{\reftab}[1]{Tab.~\ref{tab:#1}}

\newcommand{\angstrom}{\mbox{\normalfont\AA}}

\def\bfx{\boldsymbol{x}}
\def\bfy{\boldsymbol{y}}
\def\bfA{\boldsymbol{A}}

\def\bfd{\boldsymbol{d}}

\def\kms{\, {\rm km}\, {\rm s}^{-1}\, }

\newcommand{\be}{\begin{equation}}
\newcommand{\ee}{\end{equation}}
\newcommand{\ba}{\begin{eqnarray}}
\newcommand{\ea}{\end{eqnarray}}
\newcommand{\en}{\nonumber\\}

\definecolor{mlabblue}{rgb}{0, 0.4470, 0.7410}
\definecolor{mlabred}{rgb}{0.8500, 0.3250, 0.0980}
\definecolor{mlabyellow}{rgb}{0.9290, 0.6940, 0.1250}

\begin{document}


\title{Microlensing of Extremely Magnified Stars near Caustics of Galaxy Clusters}

%% Use \author, \affil, plus the \and command to format author and affiliation 
%% information.  If done correctly the peer review system will be able to
%% automatically put the author and affiliation information from the manuscript
%% and save the corresponding author the trouble of entering it by hand.
%%
%% The \affil should be used to document primary affiliations and the
%% \altaffil should be used for secondary affiliations, titles, or email.

%% Authors with the same affiliation can be grouped in a single

\author{Tejaswi Venumadhav and Liang Dai\footnote{NASA Einstein Fellow.}}
\affil{Institute for Advanced Study\\
1 Einstein Drive \\
% * <ldai@ias.edu> 2017-06-26T02:29:51.673Z:
%
% ^.
Princeton, NJ 08540, USA}

\author{Jordi Miralda-Escud\'{e}}
\affil{Instituci\'{o} Catalana de Recerca i Estudis Avan\c{c}ats\\ Barcelona, Catalonia, Spain}
\affil{Institut de Ci\`encies del Cosmos, Universitat de Barcelona \\
IEEC-UB \\
Barcelona 08028, Catalonia, Spain}


\begin{abstract}

Recent observations of lensed galaxies at cosmological distances have detected individual stars that are extremely magnified when crossing the caustics of lensing clusters. In idealized cluster lenses with smooth mass distributions, two images of a star of radius $R$ approaching a caustic brighten as $t^{-1/2}$ and reach a peak magnification $\sim 10^{6}\, (10\, R_{\odot}/R)^{1/2}$ before merging on the critical curve. We show that a mass fraction ($\kappa_\star \gtrsim \, 10^{-4.5}$) in microlenses inevitably disrupts the smooth caustic into a network of corrugated microcaustics, and produces light curves with numerous peaks. Using analytical calculations and numerical simulations, we derive the characteristic width of the network, caustic-crossing frequencies, and peak magnifications. For the lens parameters of a recent detection and a population of intracluster stars with $\kappa_\star \sim 0.01$, we find a source-plane width of $\sim 20 \, {\rm pc}$ for the caustic network, which spans $0.2 \, {\rm arcsec}$ on the image plane. A source star takes $\sim 2\times 10^4$ years to cross this width, with a total of $\sim 6 \times 10^4$ crossings, each one lasting for $\sim 5\,{\rm hr}\,(R/10\,R_\odot)$ with typical peak magnifications of $\sim 10^{4} \left( R/ 10\,R_\odot \right)^{-1/2}$. The exquisite sensitivity of caustic-crossing events to the granularity of the lens-mass distribution makes them ideal probes of dark matter components, such as compact halo objects and ultralight axion dark matter.

\end{abstract}




%%%%%%%%%%%%%%%%%%%%%%%%%%%
\section{Introduction}
\label{sec:intro}
%%%%%%%%%%%%%%%%%%%%%%%%%%%

\begin{figure}[t]
  \begin{center}
    \includegraphics[scale=0.5]{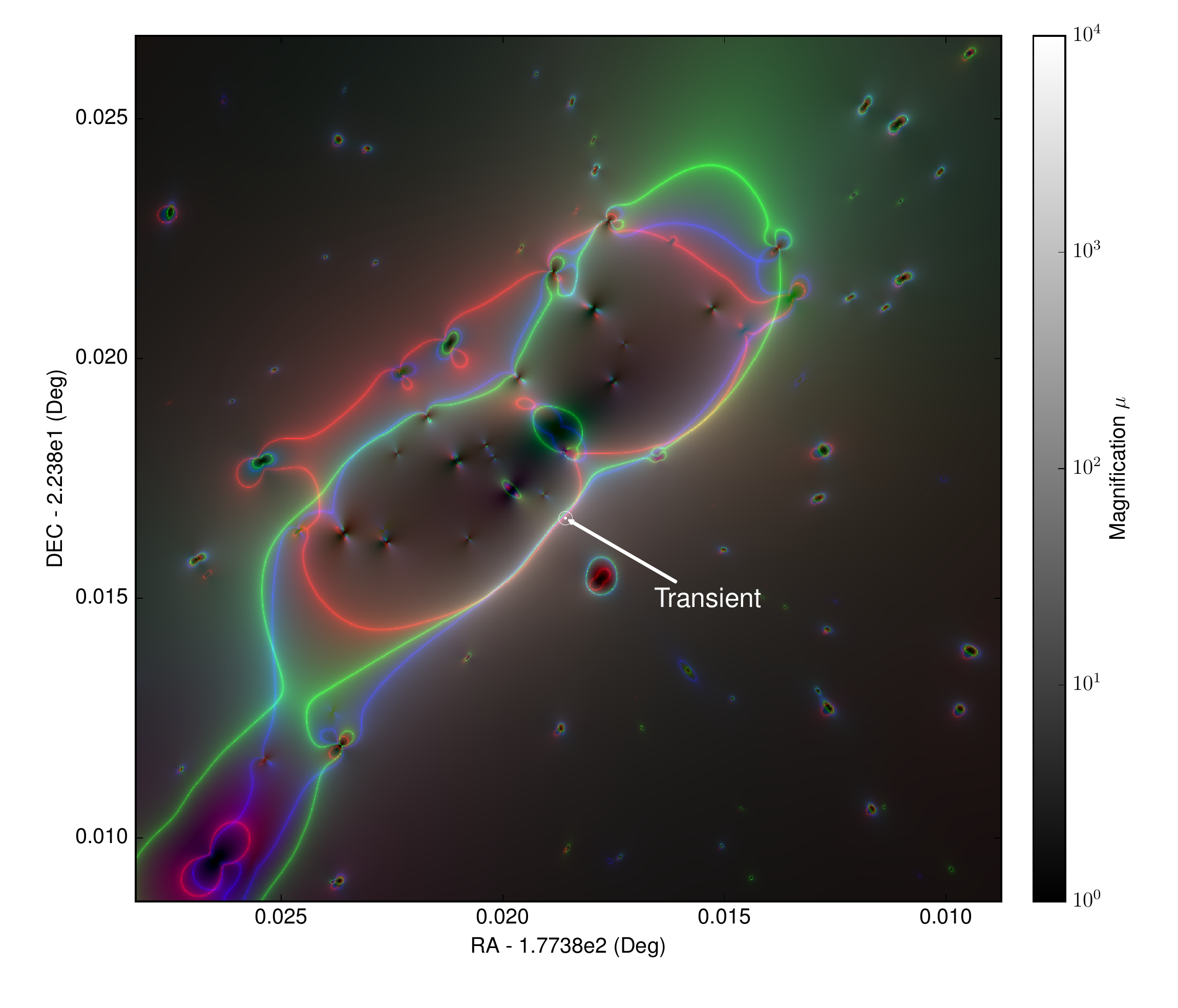}
    \caption{\label{fig:magplot} Magnification maps of the lens galaxy cluster MACS\,J1149.5+2223 for a source at $z_S=1.49$. We overplot the predictions of three lens models from the Frontier Field Lens Models project using RGB color-coding: {\tt Zitrin-ltm} \citep[red;][]{2009ApJ...703L.132Z}, {\tt Sharon} \citep[green;][]{2014ApJ...797...48J}, and {\tt GLAFIC} \citep[blue;][]{Kawamata:2015haa,2010PASJ...62.1017O}. Near the location of the transient, the three models are in good agreement on the critical curve.}
  \end{center}
\end{figure}

Galaxy clusters are the largest gravitational lenses that can make multiple 
images of, and substantially magnify, sources at cosmological distances. Clusters 
with smoothly distributed surface mass densities have critical curves with angular sizes 
of tens of arcseconds on their image planes, where the lensing magnification formally 
diverges; these curves map to caustic curves on the source plane \citep{1986ApJ...310..568B}. 
Background galaxies that lie on caustics appear as giant arcs, 
made up of two images of the part of the source inside the fold \citep[see e.g.][]{Dalal:2003kw}.
When a member star crosses from the inside of the caustic to the outside, a pair of
images approach the critical curve from both sides, brighten to a peak magnification, and 
eventually merge and disappear \citep{1991ApJ...379...94M}. Due to the extreme 
magnifications involved, single stars at high redshifts can become detectable by current 
telescopes. Recently, a candidate caustic-crossing event was reported 
in {\em Hubble Space Telescope (HST)} observations of the cluster lens MACS J1149.5+2223~\citep{2016ATel.9097....1K}. 
\reffig{magplot} shows the magnification maps of the cluster for a source at redshift $z_S = 1.49$, 
computed using publicly available lens models.

%%%%%%%%%%%%%%%%%%%%
%%%%%%%%%%%%%%%%%%%%

Three physical effects limit the peak magnification of a source during a caustic-crossing 
event. The first is the finite angular size $R$ of the source. The length of a lensed image 
is inversely proportional to the square root of the separation between the source and 
the fold caustic. Since gravitational lensing conserves the surface 
brightness, this implies that the peak magnification $\mu_{\rm peak} \sim R^{-1/2}$
\citep{1987A&A...171...49S}. The second limitation is the effect of wave diffraction. If the physical 
size of a lensed image is smaller than the geometric mean of the wavelength and 
the distance to the lens, diffraction reduces the peak magnification \citep{1983ApJ...271..551O}. 
We will see that this second limitation is unimportant in our situation of interest. The third limitation 
is the underlying granularity of the surface mass density due to the presence of discrete microlenses. 
When the source approaches the caustic of the smooth mass distribution, even a tiny mass 
fraction in microlenses substantially perturbs the lens model. The microlenses disrupt the 
macrocaustic into a network of corrugated microcaustics, whose characteristic peak 
magnifications are reduced compared to that of the smooth model 
\citep{1990LNP...360..186W, 1996IAUS..173..355B}.

An example of a population of microlenses is intracluster stars.  First postulated by 
\cite{1951PASP...63...61Z} based on observations of the Coma Cluster, the presence of 
an intracluster stellar population has been supported by the discovery of planetary nebulae
\citep{1996ApJ...472..145A,1998ApJ...492...62C}, red giant stars \citep{1998Natur.391..461F}, 
and supernovae \citep{2003AJ....125.1087G} in intracluster space. Deep exposures of 
intracluster light in typical galaxy clusters have revealed smooth halos containing as much as 
$10\%-50\%$ of all the stars in their inner regions \citep{2004ApJ...617..879L,2005MNRAS.358..949Z}.
Their origin may be explained by tidal stripping of cluster galaxies or by in-situ formation in intracluster clouds, which may themselves have been tidally ejected from galaxies \citep{Martel:2012ue,Contini:2013wha,Cooper:2014nwa}.

In this paper, we study the characteristics of microcaustic networks produced by intracluster 
stars in the vicinity of the critical curves of galaxy clusters. We derive the typical angular extent of the 
networks, both on the image and source planes. Sources crossing these networks exhibit 
numerous magnification peaks as they cross microcaustics. We derive the typical peak magnification and peak frequency within the network, and demonstrate how the source intermittently brightens and becomes visible over periods of $\sim 10^4$ years. Due to the extreme background magnification, the associated microcaustic-crossing frequency is orders of magnitude higher than that in quasar microlensing 
(\cite{1979Natur.282..561C}; see \cite{2010GReGr..42.2127S} for a detailed review).
The network of corrugated caustics, and the resultant light curves of the crossing sources, is sensitive to extremely small mass fractions of microlenses. Concomitantly, these lensing systems are sensitive to fluctuations of small amplitude in the mass density on small scales and are powerful probes of the
granularity of the mass distribution in galaxy clusters.

We therefore extend our analysis and examine the possibility that the dark matter in cluster halos 
is partly composed of compact objects, usually called MAssive Compact Halo
Objects (MACHOs), which could be primordial black holes formed in the early universe
\citep{1986ApJ...304....1P,1991ApJ...366..412G}. Constraints on the fraction of dark matter in 
MACHOs have been derived using galactic microlensing \citep{2001ApJ...550L.169A}, 
microlensing toward the Magellanic Clouds and M31 \citep{2007A&A...469..387T, 2017arXiv170102151N}, wide binaries \citep{2004ApJ...601..289C,2004ApJ...601..311Y,2009MNRAS.396L..11Q}, microlensing of the Kepler stars \citep{2011PhRvL.107w1101G, PhysRevLett.111.181302}, and stellar cluster dynamics \citep{2016ApJ...824L..31B}. In combination, these constraints rule out the possibility that MACHOs account for {\em all} the dark matter, except within a few mass windows \citep{2017arXiv170505567C}. However, the possibility remains that MACHOs comprise a small fraction of the dark matter. We discuss how this population would qualitatively modify microlensing light curves during crossings of the caustic network and show that a small fraction of dark matter in MACHOs can be probed with a sensitivity that has so far not been achieved with other existing methods. 

We begin in \refsec{analyt} by reviewing the theory of caustic crossings in smooth lens models and developing simple analytic 
estimates of the relevant scales after including microlenses. Next, we compare these 
analytic results to numerical simulations of the light curves in \refsec{numerical}, for small stellar surface mass densities, following which we extrapolate the results to the realistic surface mass densities inferred 
from existing observations. We then discuss the effect of MACHOs in \refsec{machos}. We finish 
with some concluding remarks in \refsec{concl}. We provide a technical calculation for the mean magnification through caustic crossings in \refapp{analytics}. We estimate in \refapp{application} the surface mass density in intracluster 
stars in the specific case of a candidate caustic-crossing event in the galaxy cluster MACS\,J1149.5+2223. 
Throughout this paper, we assume the {\em Planck} best-fit $\Lambda$CDM cosmology with $\Omega_m=0.308$ and
$h=0.678$~\citep{Ade:2015xua}.

%%%%%%%%%%%%%%%%%%%%%%%%%%%
\section{Structure of a corrugated caustic network: analytic estimates}
\label{sec:analyt}
%%%%%%%%%%%%%%%%%%%%%%%%%%%

% \vspace{10pt}

\subsection{Caustics in smooth lens models: definitions and notation}
\label{sec:macrolens}

% \vspace{10pt}

We first consider the vicinity of a caustic in a smooth lens model, in the absence of microlensing by point masses. In the region of geometrical optics and under the thin lens approximation, lensing is described by a mapping from the two-dimensional angular coordinates $\bf x$ on the image plane onto  the coordinates $\bt y$ on the source plane, of the form $\bt y(\bt x) = \bt x - \bm\nabla \,\psi(\bt x)$. Here, $\bm\nabla$ is the two-dimensional gradient with respect to the coordinates $\bt x$, and the projected gravitational potential $\psi(\bt x)$ is determined by the smooth mass density projected on the image plane, or surface mass density $\Sigma(\bt x)$, according to 
\citep[e.g.,][]{1986ApJ...310..568B}
\begin{align}
\label{eq:ceq}
 \nabla^2 \psi &= 2\, \frac{\Sigma(\bt x)}{\Sigma_{\rm crit}} =
 2 \kappa(\bt x) ~, \, {\rm where} \qquad
 \Sigma_{\rm crit} = \frac{c^2}{4\pi G}\, \frac{D_S}{D_L\, D_{LS}} ~.
\end{align}
Here, $\Sigma_{\rm crit}$ is the critical surface density, and in a Friedman-Robertson-Walker universe, $D_L$, $D_S$, and  $D_{LS}$ are the angular diameter distances to the lens, to the source, and from the lens to the source, respectively. The quantity $\kappa(\bt x)$ is called the convergence. Each point on the image-plane maps to a unique point on the source plane, but not vice versa---a given source can have more than one image. A smooth lens produces multiple images of sources that are within regions bounded by caustic curves on the source plane. Whenever the source crosses these caustics, multiple images annihilate or emerge in pairs on critical curves on the image plane.

The Jacobian matrix of the lens mapping takes the following form
\ba
\bt A(\bt x) \equiv
\frac{\partial\,\bt y(\bt x)}{\partial\,\bt x} =
\left( \begin{array}{cc}
1-\kappa(\bt x)-\lambda(\bt x) & -\eta(\bt x) \\
-\eta(\bt x) & 1-\kappa(\bt x)+\lambda(\bt x) \\
\end{array} \right).
\label{eq:magmatrix}
\ea
The convergence $\kappa(\bt x)$ is determined by the local surface density according to \refeq{ceq}, while the shear $\gamma(\bt x) = \sqrt{\lambda^2(\bt x) + \eta^2(\bt x)}$ is governed by the tidal force from the entire mass distribution. The inverse determinant of the Jacobian matrix is the magnification $\mu(\bt x)$ of an image at position $\bt x$, i.e., $\mu(\bt x) = 1/\det[\bt A(\bt x)]$.

The magnification formally diverges at critical curves on the image plane (which map to caustics on the source plane), so these curves are the loci of the equation $\det[\bt A(\bt x)]=0$~\citep{1986ApJ...310..568B}. They are typically located in regions where $\kappa(\bt x)$ is of order unity. We designate the characteristic angular size of the critical curves by $\theta_C$. In the simplest model of a cluster lens as a singular isothermal sphere, this angular size is the Einstein angular radius,
\ba
\theta_C = 4\pi \,\frac{\sigma^2_v}{c^2}\,\frac{D_{LS}}{D_S}  = 0.48\,
 {\rm arcmin}\, \left( \frac{\sigma}{1000\kms} \right)^2\,
   \frac{D_{LS}}{D_S} ~.
\ea
where $\sigma_v$ is the velocity dispersion of the isothermal sphere.

\begin{figure}[t]
  \begin{center}
    \includegraphics[width=15cm]{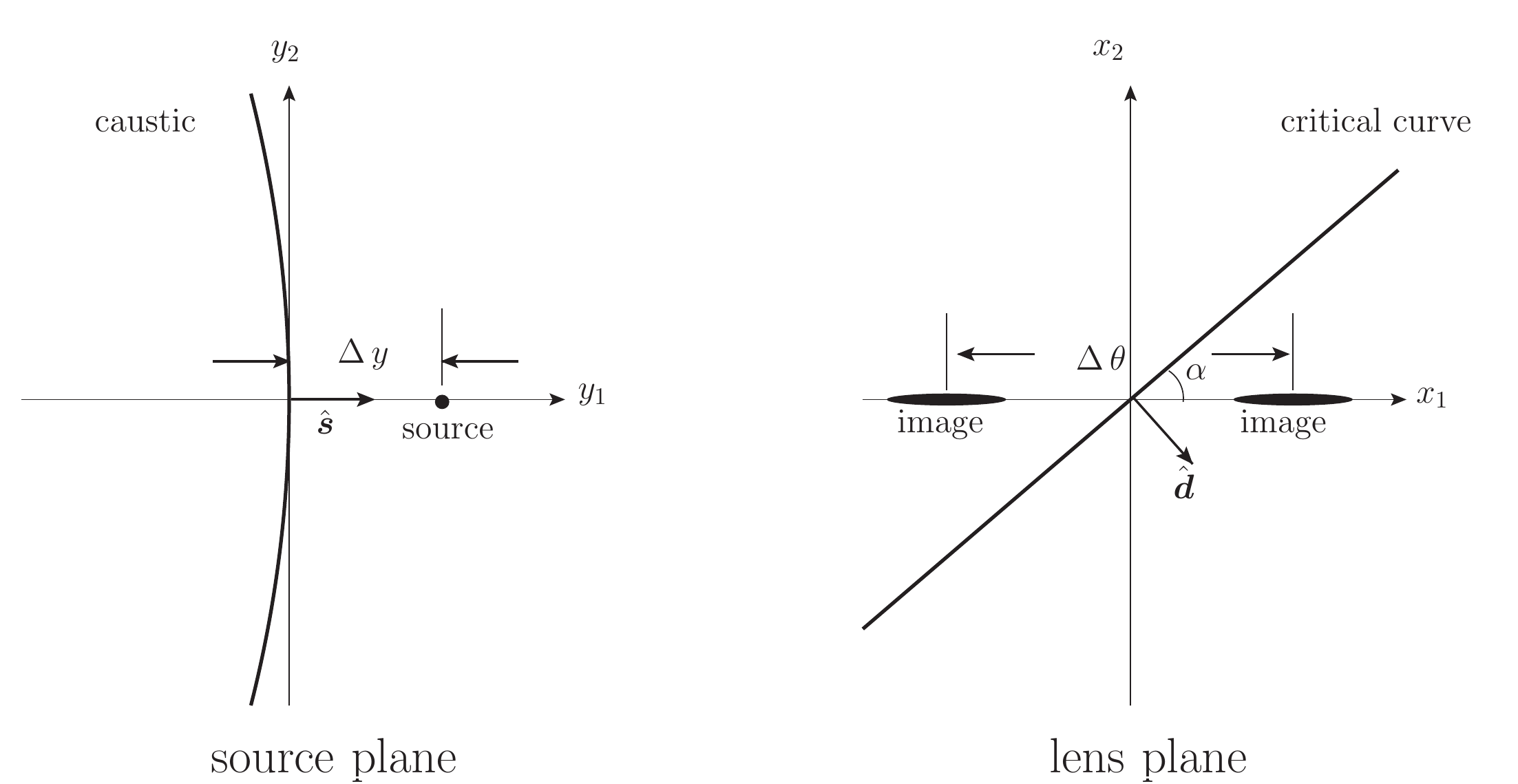}
    \caption{\label{fig:fold} Fold caustic on the source plane and the critical curve on the image plane. We choose the coordinate systems on the two planes so that their origins are on the critical curve and caustic, and they are oriented such that the horizontal axis aligns with the direction of image elongation. The critical curve is then at an angle $\alpha$ relative to this direction.}
  \end{center}
\end{figure}

The region of interest in this paper is in the immediate vicinity of a critical curve, where the image magnification is very large. It is convenient to expand the components of the Jacobian matrix $\bt A(\bt x)$ of \refeq{magmatrix} in a coordinate system centered at one chosen point on the critical curve and oriented according to the principal directions, so that $\bt A$ is diagonal and its first eigenvalue vanishes. We then have $\eta(\bt x=0)=0$ and $1-\lambda(\bt x=0)=\kappa(\bt x=0)\equiv \kappa_0$. The first-order expansion near the origin is
\ba
\kappa(\bt x) & = & \kappa_0 + \bt x \cdot (\bm\nabla \kappa)_0 ~, \\
\lambda(\bt x) & = & 1 - \kappa_0 + \bt x \cdot (\bm\nabla \lambda)_0 ~, \\
\eta(\bt x) & = &  \bt x \cdot (\bm\nabla \eta)_0 ~,
\ea
where the sub-indices ${(\cdots)}_0$ indicate quantities computed at $\bt x=0$. The inverse magnification is
\begin{align}
1/\mu(\bt x) = \det \bt A(\bt x) & = 2\,\left( 1-\kappa_0 \right)\,
(\bt d \cdot \bt x) ~, \label{eq:mud} \\
\bt d & \equiv -(\bm\nabla\kappa)_0 - (\bm\nabla \lambda)_0 ~, \label{eq:ddef}
\end{align}
where the vector ${\bf d}$ is the gradient of the eigenvalue that vanishes at the origin. To first order, the critical curve is locally given by the straight line $\bt d\cdot\bt x=0$, i.e. it is perpendicular to the vector $\bt d$. \reffig{fold} depicts this geometry: the critical curve is at an angle $\alpha$ relative to the first axis, along which images are elongated into ``arcs'' on the image plane. This angle is $\alpha = - \tan^{-1}(d_1/d_2)$, where the sub-indices $(\cdots)_1$ and $(\cdots)_2$ denote Cartesian components along the first and second axes, respectively. On the source plane, the caustic is a parabola crossing the origin and tangent to the second axis.

%%%%%%%%%%%%%%%%%%%%
%%%%%%%%%%%%%%%%%%%%

 Consider a point source at a separation $\Delta y$ from the
caustic in \reffig{fold}. When $\Delta y\, d_1 > 0$,
the source has two images of equal (absolute) magnification but opposite
parity, located symmetrically on both sides of the critical curve and
elongated along the $x_1$ direction. The two images have an angular separation
\begin{equation}
\label{eq:foldlensmap}
 \Delta \theta = 2\,x_1 = 2 \left({2
  \left|\Delta y \over d \sin{\alpha} \right| } \right)^{1/2} ~,
\end{equation}
where $d$ is the modulus of the vector $\bf d$. The total magnification of
the two images, $\mu_t$, is
\ba
\label{eq:mu1}
\mu_t = 2\, \mu = \frac{1}{2|1-\kappa_0|}
 \left( \frac{2}{\vert d\,\sin\alpha \vert \vert \Delta y \vert } \right)^{1/2}
 = \frac{2}{|(1-\kappa_0)\,d\,\sin\alpha |\, \Delta \theta} ~.
\ea 
In this equation, the quantity $(1-\kappa_0)$ is usually of order unity and the gradient $d$ is of the order $\theta_C^{-1}$. The angle $\alpha$ can be very small for a nearly spherically symmetric lens or when the fold is close to being a cusp.\footnote{A cusp is a special point along the caustic and critical line where the angle $\alpha=0$, for which a higher-order expansion is required to analyze the variation of the Jacobian matrix in its vicinity on the image plane.} In most cases, 
when images are not very close to a cusp, their magnification is of order $\mu\sim \theta_C/\Delta \theta$. 
The source size, $R$, limits the maximum magnification, which is reached when 
$\Delta y\simeq R$.

 If the source moves relative to the lens from the inside of the caustic to the outside, two of its images move toward each other along the $x_1$ axis (irrespective of the direction of the source motion, as long as it is not exactly parallel to the caustic), and gradually brighten until they merge on the critical curve and disappear. The reverse process occurs when the source moves from the outside to the inside of the caustic curve. The total magnification follows a power law with the time to image merger
\ba
\label{eq:muoft}
\mu_t(t) & = & \frac{1}{2|1-\kappa_0|}
 \left( \frac{2\,D_S}{d\,|\sin\alpha|\,v_t} \right)^{1/2}
 \frac{1}{|t-t_0|^{1/2}} \en
& = & 3.83 \times 10^6 \left( \frac{0.17}{|1-\kappa_0|} \right)
 \left( \frac{D_S}{1.7\,{\rm Gpc}} \right)^{1/2} 
 \left( \frac{5\,{\rm arcmin}^{-1}}{d\,|\sin\alpha|} \right)^{1/2}
 \left( \frac{1000\kms}{v_t} \right)^{1/2}
 \left( \frac{1\,{\rm hr}}{|t-t_0|} \right)^{1/2} ,
\ea
where $t$ is the proper time measured by the observer and $t_0$ is the time of the 
caustic crossing. Here, we use fiducial values for the distances appropriate for the 
case of the gravitational lens MACS J1149+2223, at redshift $z_L=0.54$, with a lensed 
galaxy at redshift $z_S=1.5$, as we describe in more detail in \refapp{application}
\citep[see][]{2016ATel.9097....1K}). We note that the low value of $|1-\kappa_0|=0.17$ inferred for the position of the observed transient in this cluster helps achieve a high magnification. The velocity $v_t$ is the difference in the proper velocities 
of the background source ($\bt v_s$) and the smooth lens ($\bt v_l$) with respect to 
the Earth, appropriately redshifted and projected along the direction perpendicular to the 
caustic,
\ba
\label{eq:vrel}
v_t = \left| \left( \frac{\bt v_s}{1 + z_S} - \frac{D_S}{D_L(1+z_L)} \bt v_l
 \right) \cdot  \hat{\bt s} \right|, \label{eq:projv}
\ea
where $z_S$ and $z_L$ are the source redshift and the lens redshift, respectively, and $\hat{\bt s}$ is a unit vector on the source plane perpendicular to the caustic (on the image plane, $\hat{\bt s}$ is also the direction of image elongation; it points along the first coordinate axis in the coordinate system of \reffig{fold}). With this scaling, the velocity $v_t$ measures the physical displacement at the source redshift per unit observer proper time; the fiducial value $v_t=1000 \, \kms$ reflects the typical motion of the large-scale structure.

In \refeq{muoft}, the net magnification of the pair of images formally diverges for a point source that is exactly on the caustic. As we have mentioned in the introduction, in reality, in the absence of microlenses, two effects physically limit the maximum magnification at the time of caustic crossing. They are the finite size of the source and wave diffraction. The diffraction limit is roughly \citep[see, e.g.,][]{2017PhRvD..95d4011D}
\ba
\mu_{\rm max,diff} \simeq 3\times 10^9 \,
 \left( \frac{5\,{\rm arcmin}^{-1}}{d} \right)^{2/3}
 \left( \frac{\nu}{10^{15}\,{\rm Hz}} \right)^{1/3}
 \left( \frac{0.92\,{\rm Gpc}}{D_{LS}} \right)^{1/3}
 \left( \frac{D_S}{1.7\,{\rm Gpc}} \right)^{1/3}
 \left( \frac{D_L}{1.3\,{\rm Gpc}} \right)^{1/3}
 \left( \frac{1+0.5}{1+z_L} \right)^{1/3}.
 \label{eq:mumaxdiff}
\ea
where $\nu$ is the wave frequency and $z_L$ is the lens redshift. If we ignore the wave effects, a uniform disk source with physical radius $R$ reaches a maximum magnification \citep{1991ApJ...379...94M}
\ba
\label{eq:mumax-smooth}
\mu_{\rm t,max} = \frac{1.4}{|1-\kappa_0|}\,
\left( \frac{D_S}{2\, R\,d\,|\sin\alpha|} \right)^{1/2} \simeq
4 \times 10^6\, \left( \frac{0.17}{|1-\kappa_0|} \right)
\left( \frac{D_S}{1.7\,{\rm Gpc}} \right)^{1/2}
\left( \frac{5\,{\rm arcmin}^{-1}}{d\,|\sin\alpha|} \right)^{1/2}
\left( \frac{10\,R_\odot}{R} \right)^{1/2},
\ea
where $R_\odot = 6.96 \times 10^5\,{\rm km}$ is the solar radius. For observations of cluster lenses at cosmological distances at optical or near-infrared wavelengths, $\mu_{\rm t,max} \ll \mu_{\rm max,diff}$, and thus diffraction effects are unimportant even for stars as small as white dwarfs. The duration of the magnification peak (in the observer frame) is roughly the time it takes for the stellar disk to traverse the caustic,
\ba
\label{eq:timescale-peak}
\tau_{\rm peak} = \frac{2R}{v_t} \simeq 4\,{\rm hr}\,
\left(\frac{R}{10\, R_\odot}\right) \left( \frac{1000\kms}{v_t} \right) .
\ea
Since a star of solar luminosity at redshift $z_S \simeq 1.5$ has an apparent magnitude $m \sim 50$, a star with $L\sim 10^3\, L_{\odot}$ magnified by a factor of $\mu_{\rm t} \sim 10^6$ would become visible to {\em HST} at magnitude $m \sim 27.5$ during one of these caustic crossings. However, as we shall show next, the caustic-crossing peak magnification is actually much lower than this simple estimate due to the presence of microlenses in the intracluster region.

%%%%%%%%%%%%%%%%%%%%%%%%%%%
\subsection{Caustic crossing with microlenses}
\label{sec:microlens}
%%%%%%%%%%%%%%%%%%%%%%%%%%%

In this section, we use simple analytical scaling arguments to demonstrate how point masses, such as stars, in the vicinity of a macroscopic lens model critical curve transform the macrocaustic curves (with only a few cusps) into a network, or band, of corrugated microcaustics. We show how this systematically reduces the peak magnification, so that stars of higher luminosity, $L \sim 10^5\,L_\odot$, are needed to make caustic-crossing events visible to current telescopes at magnitude $27.5$. At the same time, each star crosses a very large number of microcaustics as it moves relative to the foreground lens across the network, so that the rate at which caustic crossings occur is greatly enhanced. We estimate the characteristic mean and peak magnifications and time-scales of the resulting light curves.

We model a cluster lens as a smooth mass distribution (in the sense of coarse-graining) with critical curves of characteristic angular size $\theta_C$, and a total projected mass $M_C$ within, but with part of this mass actually in the form of point masses. In \refapp{application}, we estimate the contribution of intracluster stars to the convergence near the critical curves to be $\kappa_\star\sim 0.01$, for the cluster lens shown in \reffig{magplot}.

It is worth pausing to clarify an issue of nomenclature in order to avoid confusion in what follows. In order to numerically study the impact of microlenses, we restrict them to a circular region of radius $\mathcal{R}$ on the image plane, within which we distribute them randomly (the size $\mathcal{R}$ is large enough to ignore the influence of microlenses outside the region, except near the edges that are not used). A given lens model in the literature has a surface mass density $\kappa(\bt x)$ due to an entirely smooth component. We refer to this as ``the macroscopic lens model,'' or simply ``the macroscopic model,'' and we use the subscript ${}_{\rm M}$ for the associated quantities. When including point masses with an average surface mass density $\kappa_*$ within $\mathcal{R}$, we also appropriately reduce the smooth component to preserve the total mass. We use the term ``the smooth lens model,'' or simply ``the smooth model,'' and the subscript ${}_{\rm S}$, to refer to the smooth component of the model with a surface mass density $\kappa_{\rm S}(\bt x) = \kappa(\bt x) - \kappa_\star$, without any microlens. This is distinct from the model we start with, i.e., the macroscopic lens model with the fully smooth surface density $\kappa(\bt x)$. Both models have the same shear within the region $\mathcal{R}$. However, they have different critical curves on the image plane---we use the term ``the smooth critical curve'' for the smooth lens model and the term ``the macrocritical curve'' for the macroscopic lens model. We refer to the caustic of the latter lens model as the ``macrocaustic''.

\subsubsection{The band of corrugated microcritical curves}
\label{sec:critanalytic}

Consider a point microlens with mass $M_\star \ll M_C$, located close to the smooth critical curve on the image plane. We use the same notation as in \reffig{fold} (which we take to represent the smooth model), in which the coordinate origin lies on the smooth critical curve that is inclined at an angle $\alpha$ with respect to the axis of image elongation $\hat{\bt s} = \hat{\bt x}_1$, and the coordinate $\bt x$ represents the angular displacement from this curve.

\begin{figure*}[t]
\begin{center}
    \includegraphics[width=12cm]{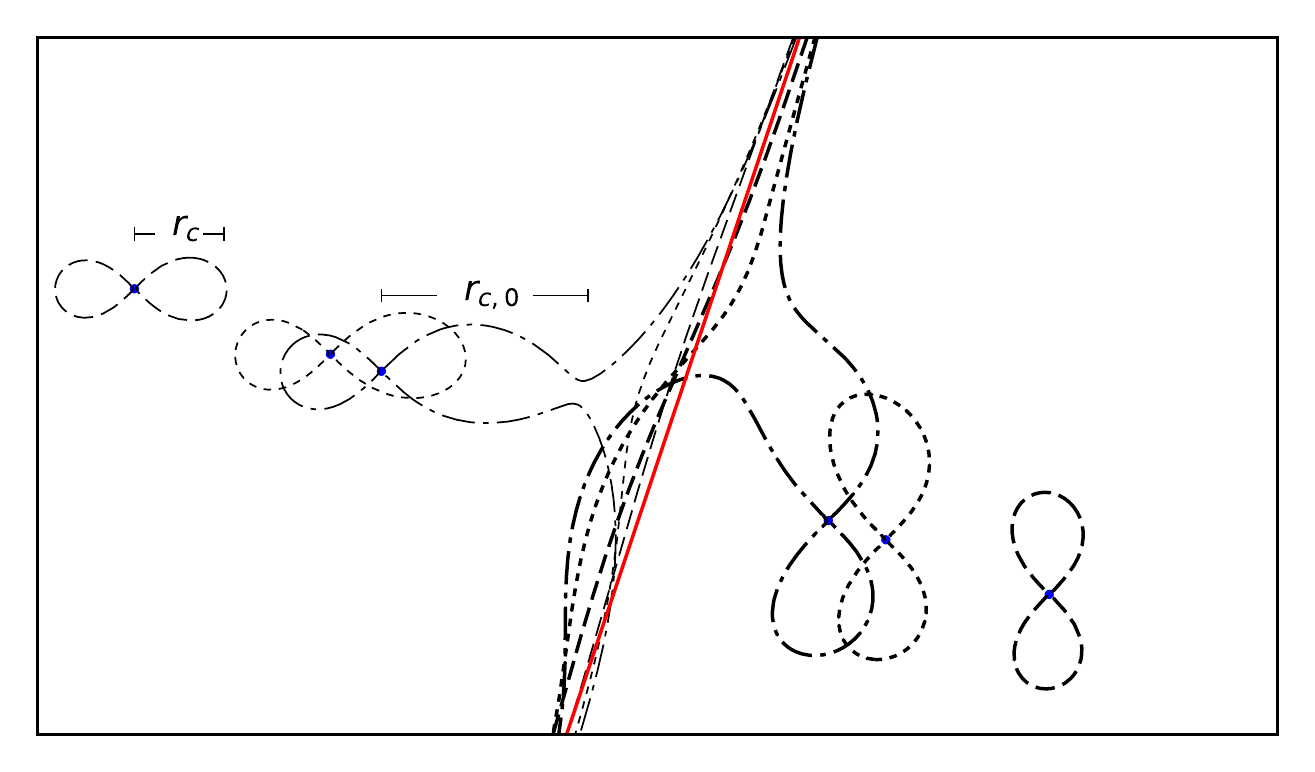}
\caption{\label{fig:critical_zoomed} 
Solid red line: critical curve of the smooth lens model.
Dashed, dotted and dashed-dotted lines: critical curves perturbed by a single
microlens (blue dot) as it gradually approaches the smooth-model critical curve. Thick or thin lines are used when the point mass
is on the right or the left side of the smooth critical curve, respectively.
The size $r_c$ of the branch of the critical curves around the microlens
gradually increases as the smooth critical curve is approached, up to a 
value $r_{c,0}$, where it merges with the perturbed smooth critical curve.}
\end{center}
\end{figure*}

In the absence of the smooth lens, the characteristic size of the microcritical curve is the point-mass Einstein radius
\begin{align}
  \theta_\star & = \left( \frac{4\,G\,M_\star}{c^2}\, \frac{D_{LS}}{D_L\,D_S} \right)^{1/2} = 1 \ \mu{\rm as} \left( \frac{M_\star}{0.3 \, M_{\odot}} \right)^{1/2} \left( 2.55 \, {\rm Gpc} \frac{D_{LS}}{D_L\,D_S} \right)^{1/2}.
  \label{eq:einsteinpt}
\end{align}
We first study the case of a small surface density in point masses, so that we can neglect the influence of nearby microlenses on microcritical curves near a point mass. In this case, we can approximate the Jacobian matrix as the sum of the contributions of the smoothly distributed mass, and a single point mass. We can make two further approximations in the vicinity of the smooth critical curve: first, the eigenvalue $1-\kappa_{\rm S}-\lambda_{\rm S} = \bt d \cdot {\bf x} \ll 1$ for the smooth lens model, which holds when the separation $|{\bf x}| \ll \theta_C$; second, the microcritical curve around the point mass is much larger in size than the Einstein radius $\theta_*$ of \refeq{einsteinpt}. We will see that the latter results from the large magnification of the smooth lens model in the vicinity of its critical curve. Consequently, the separation between the microcritical curve and the point mass is typically much larger than the angular scale $\theta_\star$. Under these two assumptions, we can neglect the shear component $\eta$ due to the point mass and compute the magnification using $\mu^{-1} = \det \bfA \simeq 2\,(1-\kappa_{\rm S, 0})\,(1-\kappa_{\rm S}-\lambda)$, where the shear $\lambda$ includes the contribution of the point mass. The critical curve satisfies
\ba
\label{eq:cc-pt}
1 - \kappa_{\rm S} - \lambda_{\rm S} - \lambda_{\star} = 
\bt d \cdot \bt x + \theta^2_\star\,
\frac{(x_1 - x_{\star,1})^2 - (x_2 - x_{\star,2})^2}
 {\left| \bt x - \bt x_\star \right|^4} = 0.
\ea 
The solution to \refeq{cc-pt} consists of two branches. One branch is the original smooth critical curve ($\bt d \cdot \bt x=0$), with a perturbation that is largest at the closest distance to the point mass. The second, dumbbell-shaped, branch is in the vicinity of the point mass; the dumbbell aligns with the principal axes of the shear defined by the smooth lens and changes orientation for point masses inside or outside the smooth critical curve. This can be seen in \reffig{critical_zoomed}, which shows critical curves that have been calculated exactly when a point mass is located at various distances to the smooth critical curve and are well-approximated by \refeq{cc-pt}.

%---------------------------------------------------------
%---------------------------------------------------------

The dumbbell-shaped critical curves have a characteristic size
\be
\label{eq:rc}
 r_c \sim \theta_\star |\bt d\cdot\bt x|^{-1/2} \gg \theta_\star ~. 
\ee
Closed critical curves near microlenses are generic when external shear breaks circular symmetry~\citep{1979Natur.282..561C, schneider1999gravitational}. When the magnification of the smooth model is very large, the area of the closed critical curve, and hence the microlensing cross-section, is greatly increased. Note that these curves never pass through the point mass, and they are actually closed and differentiable (\refeq{cc-pt} is no longer valid very near the point mass), even though this is not apparent at the resolution of \reffig{critical_zoomed}.

 When the point mass is close enough to the critical curve, i.e.,
when the separation $|\bt x| \simeq r_c$, the two branches merge. We
solve for this separation by substituting in \refeq{rc}:
\ba
|\bt x| = r_{c0} \simeq \left( \frac{\theta^2_\star}{d} \right)^{1/3}
 = \theta_\star \left( \theta_\star d \right)^{-1/3}.
\ea
\reffig{critical_zoomed} shows the behavior of both branches of the perturbed critical curve as the microlens is brought successively closer to the smooth critical curve.

This characteristic separation, at which critical curves around a single point mass merge with the ones in the smooth lens model, defines a threshold surface mass density $\kappa_c$:
\ba
\label{eq:kappac}
\kappa_c \simeq \left( \frac{\theta_\star}{r_{c,0}} \right)^2 =
 \left( \theta_\star\, d \right)^{2/3} \simeq
 1.9 \times 10^{-5}
 \left( \frac{\theta_\star}{\mu{\rm as}} \right)^{2/3}
 \left( d \cdot 12\, {\rm arcsec} \right)^{2/3} \ll 1.
\ea
Physically, at this surface mass density, a large portion of the smooth critical curve is strongly perturbed and merges with
the ones around nearby point masses, which in turn strongly interact with each other in this region. The threshold value $\kappa_c$ increases with the microlens mass as $M_\star^{1/3}$.

If the surface mass density in point masses is below the threshold value, i.e., $\kappa_\star \lesssim \kappa_c$, the smooth critical curve is not highly perturbed except over a small fraction of its length. 
We see from \refeq{kappac} that the value of $\kappa_c$ is numerically small for the fiducial mass scale $M_\star$ associated with microlensing by ordinary stars in a typical cluster lens. Thus, reasonable densities of intracluster stars can easily satisfy $\kappa_\star \gg \kappa_c$. In this case, the smooth critical curve is strongly perturbed everywhere and forms a network due to the combined shear of many microlenses. We can analytically estimate the general properties of this network, or band, of corrugated microcritical curves. For simplicity, we assume that (a) all microlenses have the same mass $M_\star$, and therefore identical angular Einstein radius $\theta_\star$, and (b) the average surface mass density in point masses $\kappa_\star$ is uniform in the vicinity of the smooth critical curve (up to Poissonian fluctuations). As noted earlier, the smooth component contributes the remaining surface mass density, $\kappa(\bt x) - \kappa_\star$. 

 The typical separation between neighboring point masses is
\be
\label{eq:rs}
r_s \simeq \theta_\star \, \kappa^{-1/2}_\star ~. 
\ee 
Near the smooth critical curve, the microcritical curves of point masses merge with those of the neighboring masses, and form a band extending out to a distance $r_w$. We can estimate this distance by equating the microcritical curve size $r_c$ from \refeq{rc} to the interlens separation from \refeq{rs}. This yields the estimate
\ba
\label{eq:rw}
r_w \simeq \frac{\kappa_\star}{d} \simeq
 {1\over d} \left( d\, \theta_\star \right)^{2/3} 
\left( \frac{\kappa_\star}{\kappa_c} \right) 
\simeq \theta_\star\, \kappa_c^{-1/2} \,
\left( \frac{\kappa_\star}{\kappa_c} \right)~.
\ea
This length scale has a simple physical interpretation: it is the scale at which the eigenvalue $1 - \kappa - \lambda$ (which determines the magnification) typically receives comparable contributions from the smooth component and from the microlenses. %over most of the length of critical curves.

\begin{figure*}[t]
\begin{center}
    \includegraphics[width=0.9\textwidth]{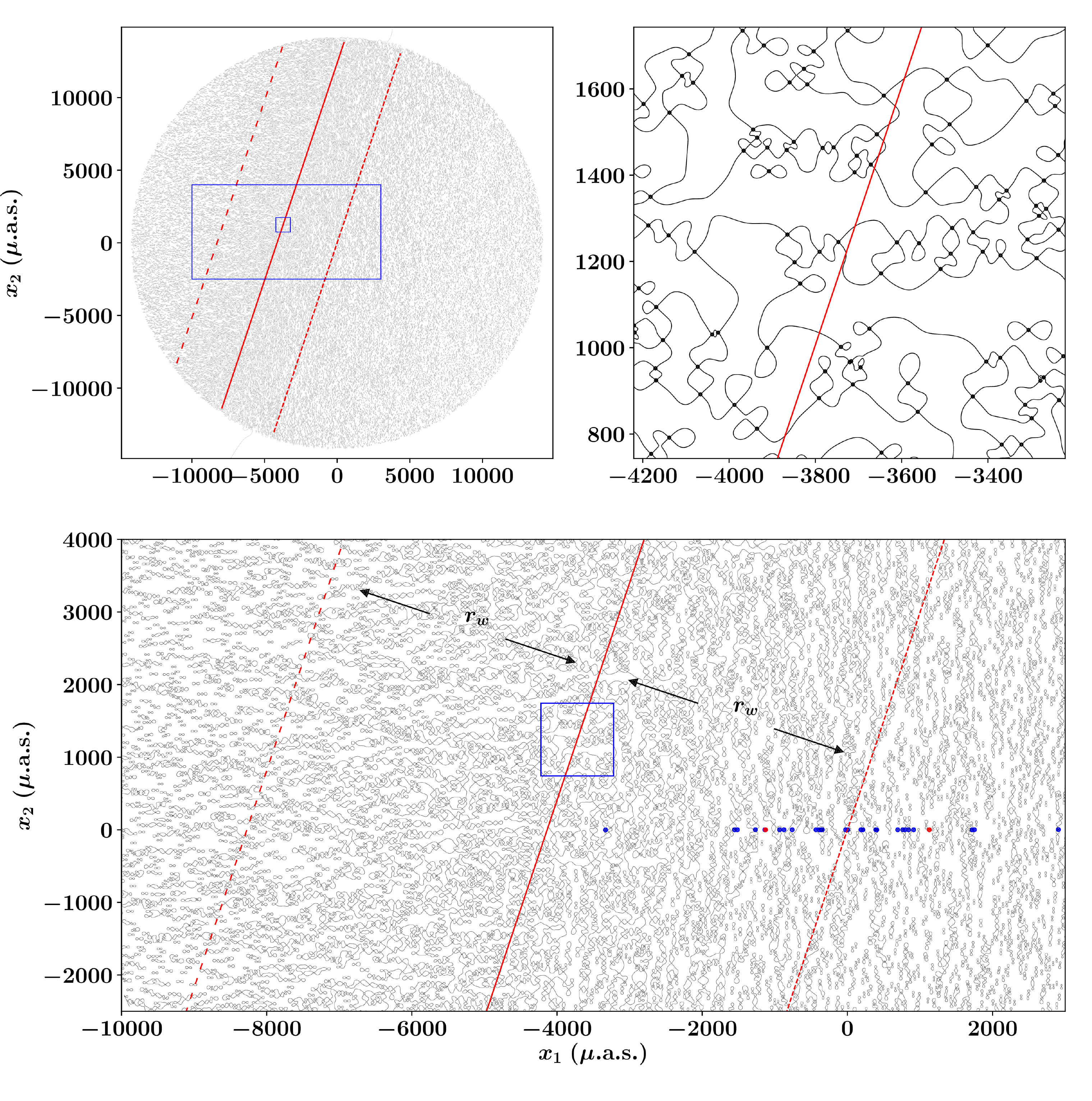}
    \caption{\label{fig:critical_network} Critical curves for a cluster lens model with $d = 4.97\, {\rm arcmin}^{-1}$, and microlenses with $\theta_\star = 1 \, \mu \, {\rm as}$ and surface mass density $\kappa_* = 17.1 \, \kappa_{\rm c} = 3.25 \times 10^{-4}$. The microlenses are included in the numerical calculation out to a radius $\mathcal{R} = 1.4 \times 10^4\, \theta_\star$. Starting with the top-left panel and moving counterclockwise, the panels show details on successively finer scales---{\color{blue} blue} rectangles mark regions that are zoomed into. Short- and long-dashed {\color{red} red} lines mark the edges of the band within which microcritical curves around individual masses join to form a network, according to our analytic approximations. The solid and short-dashed {\color{red} red} lines are critical curves for the smooth and macroscopic lens models, respectively (the former has its surface mass density reduced by $\kappa_\star$, but without adding any microlenses). The {\color{blue} blue} circles in the bottom panel mark the magnified micro-images ($\mu > 1$) of a point source at a separation $\Delta y = 0.05 \ \mu.{\rm as}$ from the macrocaustic on the source plane, and are distributed within elongated regions around the macro-images, shown by {\color{red} red} circles. In the top-right panel, black dots mark the positions of the point masses. Note that at sufficiently high resolution the critical lines never actually ``touch'' the point masses nor intersect one another.}
\end{center}
\end{figure*}

\reffig{critical_network} shows an example of critical curves on the image plane, and the thickness $r_w$ of the critical curve network, for a population of microlenses with $\kappa_* \simeq 17 \, \kappa_c$. The underlying macroscopic model is that of \reffig{magplot} at the location of the transient, but brought to the orientation of \reffig{fold} with an extra rotation. The critical curves shown are exactly calculated as contours of the condition $\mu^{-1}(\bt x) = 0$. The solid line shows the smooth critical curve, and the dashed lines show the edges of the network, as discussed above. Note that one edge of the band (the short-dashed line in the figure) is the critical curve of the macroscopic lens model, i.e., the model without microlenses and with the convergence contribution $\kappa_\star$ not removed from the region $\mathcal{R}$. Note also that the figure is centered on this macrocritical curve instead of on the one for the smooth lens model (the solid line). 

We observe that outside the band of total width $2\,r_w$, the critical curves typically form loops associated with only one or a few point masses, while inside the band the loops are either larger or join the corrugated network of critical curves that separates regions with two different signs of the eigenvalue $1-\kappa-\gamma$. This figure confirms our expectation, from our analytic derivation, for the width of the band of corrugated critical curves. 

%%%%%%%%%%%%%%%%%%%%%
%%%%%%%%%%%%%%%%%%%%%

\subsubsection{The band of corrugated microcaustics: total width and peak magnifications}
\label{sec:caustics}

We now consider the statistics of microcaustic crossings on the source plane. Throughout, we use the coordinate system of \reffig{critical_network}, with the origin on the {\em macroscopic} critical curve (and not the one for the smooth model).

We first compute the width, $s_w$, of the band of corrugated caustics on the source plane. This band is the result of mapping the network of microcritical curves of \reffig{critical_network} onto the source plane. We assume that the stochastic deflection from all point masses is small compared to the width $s_w$; we will check the validity of this assumption later. We can then use the lens map of the macroscopic lens model to relate the widths of the networks on the image and source planes. We therefore use \refeq{foldlensmap} to find the source displacement $\Delta y = s_w$ needed to create an image at position $x_1=2\,r_w$ at the edge of the corrugated band, as seen in \reffig{critical_network}), to obtain
\begin{align}
\label{eq:sw}
s_w = \frac12\, d\, |\sin\alpha |\,
 \left( \frac{2\,r_w}{|\sin\alpha |} \right)^2 \simeq
 \frac{2\, \theta_\star}{|\sin\alpha|} \, \kappa_c^{1/2} \,
 \left( \frac{\kappa_\star}{\kappa_c} \right)^2 \, ,
\end{align}
where the factor $|\sin\alpha|$ dividing $2\,r_w$ accounts for the fact that the band of corrugated critical curves is inclined at an angle $\alpha$ to the direction $\hat{\bt s}$ of large magnification (the horizontal direction in \reffig{critical_network}).

\begin{figure}[t]
\begin{center}
    \includegraphics[width=0.8\textwidth]{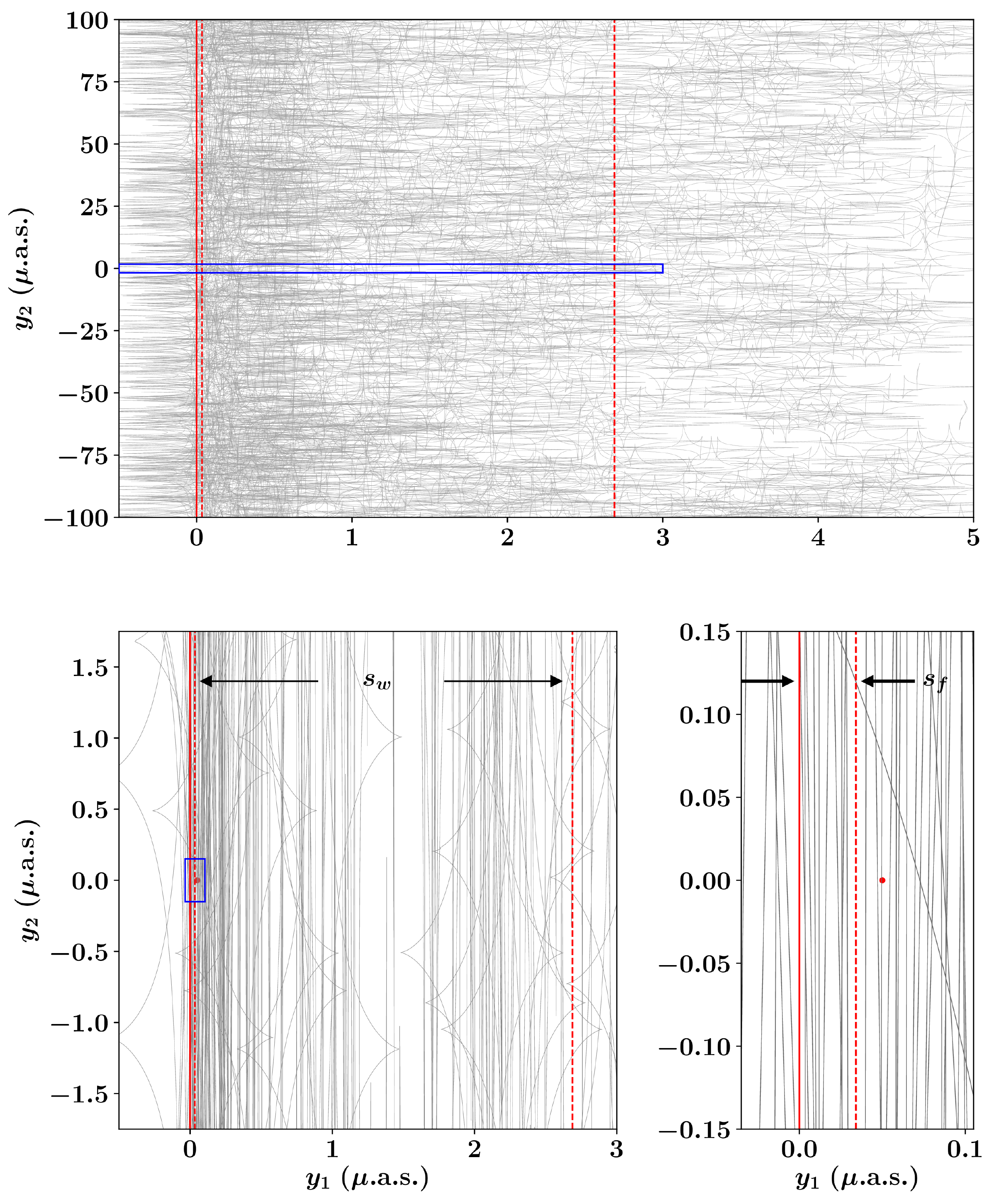}
\caption{\label{fig:caustics_network} Caustics for a cluster lens model with microlenses added: the parameters are identical to those of \reffig{critical_network}, i.e., $\kappa_\star = 17.1\, \kappa_{\rm c} = 3.25 \times 10^{-4}$. Starting with the top panel and moving counterclockwise, the panels show details on successively finer scales---{\color{blue} blue} rectangles mark regions that are zoomed into. The solid {\color{red} red} line is the caustic of the macroscopic lens model. Microcaustics join to form a corrugated network within the band of width $\simeq s_w$. These caustics are not distributed uniformly---their density increases toward the macrocaustic, and peaks in a region of width $\simeq s_f$ (shown in the lower-right panel), where the frequency of crossings maximizes. The {\color{red} red} circle marks the position of a point source whose images were shown in \reffig{critical_network}.}
\end{center}
\end{figure}

\reffig{caustics_network} shows the microcaustic network on the source plane for the lens map of \reffig{critical_network}. The caustics are exactly obtained by mapping the critical curves to the source plane under the lens map. The source generating the blue images in \reffig{critical_network}) is the red circle in the lower panels. The rightmost dashed red line defines the boundary of the caustic network of width $s_w$.

%%%%%%%%%%%%%%%%%%%%%
%%%%%%%%%%%%%%%%%%%%%

A source traversing this band will show many magnification peaks, corresponding to the crossing of multiple microcaustics. This entire set of peaks occur over a time
\ba
\label{eq:tauw}
\tau_{w} \simeq \frac{s_w\,D_S}{v_t} \simeq
 \frac{26 \,{\rm days}}{|\sin\alpha|}\,
 \left( \frac{\kappa_\star}{\kappa_c} \right)^2
 \left( \frac{\theta_\star}{\mu{\rm as}} \right)^{4/3}
 \left( d\cdot 12 {\rm arcsec} \right)^{1/3}
 \left( \frac{D_S}{1.7\,{\rm Gpc}} \right)
 \left( \frac{1000\,\kms}{v_t} \right) \, .
\ea
We can estimate the total number of microcaustic crossings over the time $\tau_w$ by assuming that there is typically one critical curve over a distance equal to the mean separation between neighboring point masses, $r_s$ (as defined in \refeq{rs}). The justification is that whenever we move over a length $r_s$ on the image plane, the point-mass contribution to the shear changes substantially, and new roots of the equation $\mu^{-1} = \det \bt{A} = 0$ are likely to be found, as long as we are within the width $r_w$ (as defined in \refeq{rw}).\footnote{A rigorous proof would need to consider the joint distributions of deflection and shear near the macrocaustic; we do not attempt to provide it in this paper. For related work in the literature, see, e.g., \cite{1994A&A...288...19S}.} The total number of microcaustics is therefore
\ba
\label{eq:Nc}
N_c \simeq \frac{2\,r_w}{r_s\,|\sin\alpha|} \simeq
 \frac{2}{|\sin\alpha|}\, \frac{\kappa^{3/2}_\star}{\kappa_c}\,
 \left( d\, \theta_\star \right)^{-1/3} =
 \frac{2}{|\sin\alpha|}\,
 \left( \frac{\kappa_\star}{\kappa_c} \right)^{3/2} \, .
\ea
As earlier, the factor $1/|\sin\alpha |$ accounts for the fact that as the source crosses the microcaustic network, its images move at an angle $\alpha$ relative to the macrocritical curve.

Next, we estimate the typical peak magnifications achieved during crossings of individual microcaustics. For sources of typical stellar radii, we can treat the lens model as an isolated fold in the vicinity of each crossing (we will justify this later). Hence, the peak magnification of a source of radius $R$ is given by \refeq{mumax-smooth}, with the gradient $d$ replaced by the local value $d_\star$ of the total eigenvalue $1-\kappa-\lambda$ (including the effects of all point masses) evaluated at the point where the two highly magnified micro-images merge on the microcritical curve. When $\kappa_\star \gtrsim \kappa_c$, the closest microlenses dominate the gradient $d_\star$. If the source is outside the band of width $s_w$, the microcritical curves are typically at a distance $r_c$ (from \refeq{rc}) from a point mass. If the source is instead within the width $s_w$, the microcritical curves are typically separated by the mean distance between the point masses, $r_s$ (\refeq{rs}). The modulus of the eigenvalue gradient is, for these two cases, respectively,
\ba
  d_\star 
  \sim \begin{cases}
  \theta_\star^2/r^3_s \simeq \kappa^{3/2}_\star/\theta_\star \simeq d \left( \kappa_*/\kappa_c \right)^{3/2}, 
  & \vert \bfy \vert < s_w, \\  
  \theta_\star^2/r^3_c \simeq \theta_\star^{-1}\, | \bfd \cdot \bfx |^{3/2}
 \simeq d \left( \kappa_*/\kappa_c \right)^{3/2}
 \left( | \bfy |/s_w \right)^{3/4}, & | \bfy | > s_w.
  \end{cases}
\label{eq:dstar}
\ea
In the second equation, we used the fold solution to map between the image and the source planes. We see that when the surface mass density $\kappa_\star \simeq \kappa_c$, the typical eigenvalue gradient $d_\star \simeq d$. As we increase $\kappa_\star$, we find from Equations \eqref{eq:Nc} and \eqref{eq:dstar} that both the gradient $d_\star$ and the total number of microcaustic crossings $N_c$ increase as $\kappa_\star^{3/2}$. Substituting \refeq{dstar} into \refeq{mumax-smooth}, the typical peak magnification at each microcaustic crossing is, for $\kappa_\star \gtrsim \kappa_c$:
\ba
\label{eq:microlen_mumax}
\mu_{\rm peak} \simeq \frac{1.4}{|1-\kappa_0|}
 \left( \frac{D_S}{2\,R\,d_\star | \sin \alpha_\star |} \right)^{1/2}
 \simeq \frac{1}{|1-\kappa_0|} \left( \frac{D_S}{R\,d} \right)^{1/2}
 \left( \frac{\kappa_c}{\kappa_*} \right)^{3/4} \times 
 \begin{cases}
   1, & \vert \bfy \vert < s_w, \\
   \left( s_w/ |\bfy |  \right)^{3/8}, & | \bfy | > s_w.
 \end{cases}
\ea
Here, $\alpha_\star$ is the angle between the local principal axis $\hat{\bt s}_\star$ and the microcritical curve at the point where two images merge. We have eliminated in the last expression the numerical factor $1.4/(2\, | \sin\alpha_\star | )^{1/2}$, which varies for each caustic crossing but is typically of order unity. 

From \refeq{microlen_mumax}, we see that the characteristic peak magnification at microcaustic crossings is reduced relative to that of the macroscopic lens model by the factor $(\kappa_c/\kappa_\star)^{3/4}$. Furthermore, it remains roughly constant
within the width of the band of corrugated microcaustics, i.e., over the timescale $\tau_w$. Outside the band, the microcritical lines are typically closer to the point masses, and the peak magnifications decrease. The time $\tau_w$ is very long in realistic situations: for example, for $\kappa_\star/\kappa_c \simeq 10^3$ and the fiducial values of the source velocity $v_t$ and other variables in \refeq{tauw}, $\tau_w$ can be as long as $\sim 10^5$ years, during which the peak magnification for microcaustic crossings remains roughly constant. During this long period, highly magnified images sporadically appear within the image-plane band of width $2\, r_w \simeq 2\, \kappa_\star/d$ (see \refeq{rw}). If the strength of intracluster light implies $\kappa_\star \simeq 10^{-2}$, and $1/d \simeq 12\, {\rm arcsec}$, the width $2\, r_w\simeq 0.2\, {\rm arcsec}$ will be resolvable by a space telescope.

\subsubsection{The frequency of microcaustic crossings}
\label{sec:freqc}

In the previous subsection, we derived the timescale $\tau_w$ over which a source crosses the band of corrugated microcaustics, and the total number $N_c$ of microcaustic crossings. Crossings during this period therefore have an average rate $N_c/\tau_w$, but they are not uniformly distributed over the entire duration. The microcritical curves are approximately uniformly distributed within the width $2\,r_w$ on the image plane (see \reffig{critical_network}), but they are mapped to the source plane with a density that is proportional to the inverse square root of the separation from the macrocaustic. This density does not increase arbitrarily, but rather hits a natural limit determined by the stochastic deflection due to the microlenses. To better understand this, we briefly discuss the probability distribution of the deflection angle $\bm{\alpha}_{\rm ml}$ due to a random distribution of point masses, which was derived by \cite{1986ApJ...306....2K}.

Consider a set of $N_{\rm ml}$ point masses acting as microlenses with Einstein radius $\theta_\star$ and surface density $\kappa_\star$. The characteristic deflection is the one at the typical separation to the nearest point mass, $r_s\simeq \theta_\star\, \kappa_\star^{-1/2}$, which equals $\theta_\star\, \kappa_\star^{1/2}$. A detailed calculation of the distribution of the total deflection by all point masses, $p(\bm{\alpha}_{\rm ml})$, shows that it has a Gaussian core with standard deviation
\begin{align}
\label{eq:sf}
  s_f & = \left[ \ln{\left( 3.05\, N_{\rm ml}^{1/2} \right)} \right]^{1/2}\, \theta_\star\, \kappa_\star^{1/2}
 \equiv \mathcal{C}_{\star}\, \theta_\star\, \kappa_\star^{1/2},
\end{align}
where the last equality defines a quantity $\mathcal{C}_{\star}$ that weakly depends on the number of microlenses $N_{\rm ml}$ in the region of interest: it originates in a Coulomb logarithm term in the cumulative deflection due to distant microlenses. In numerical estimates, the appropriate value of $N_{\rm ml}$ is the number of the microlenses whose contribution to the deflection varies significantly over the distribution of images. We derive the width of the image distribution later in \refsec{magc}: the resulting number $N_{\rm ml}$ gives an estimate of $\mathcal{C}_\star \simeq 2.5$ for the surface mass density $\kappa_\star = 0.01$. 

The Gaussian shape of the core is a consequence of the central limit theorem, which applies when the deflection is not dominated by one or a few nearby point masses. At large values of the deflection, the Gaussian core switches to a power-law tail that is governed by the chance of being close to a single point mass. This power-law tail only affects images of low magnifications, so we neglect it for the rest of this discussion.

The typical random angular deflection, $s_f$, is the limit beyond which we cannot use the macroscopic lens model to map between the image and source planes. This random deflection has two aspects relevant to our calculation: (a) for a point source, it smooths the mean magnification (averaged over many random realizations of the microlenses) away from its value in the macroscopic lens model (\refapp{analytics} presents a detailed derivation of the resulting mean magnification curve; see \reffig{meanmu}), and (b) it smooths the distribution of microcaustics and hence limits their maximum density, which is reached at a distance $\simeq s_f$ from the macrocaustic.

From the discussion at the beginning of this section, the microcaustic density on the source plane varies as $dN_c/dy_1 \propto y_1^{-1/2}$ within the band of corrugated caustics when the separation $| y_1 | > s_f$. Therefore, the number of microcaustics within a source-plane width $s_f$ of the macrocaustic is approximately
\begin{equation}
\label{eq:ncf}
 N_{cf} \simeq N_c \left( s_f\over s_w \right)^{1/2} \simeq
   \left( 2\, \mathcal{C}_{\star} \over | \sin\alpha | \right)^{1/2}
  \left( \kappa_\star \over \kappa_c \right)^{3/4} ~.
\end{equation}
The source crosses this width in a time
\begin{align}
  \label{eq:tauf}
  \tau_{f} \simeq \frac{s_f\,D_S}{v_t} =
    32 \, {\rm days}
  \left( \frac{\theta_\star}{\mu{\rm as}} \right)^{4/3}
  \left( {d \cdot 12 \, {\rm arcsec}} \right)^{1/3}
  \left( \frac{\kappa_\star}{\kappa_c} \right)^{1/2}\, \left( {\mathcal{C}_{\star}\over 2.5} \right)\,
   \left( \frac{D_S}{1.7\,{\rm Gpc}} \right) \, \left( \frac{1000\,\kms}{v_t} \right)\,.
\end{align}
The maximum microcaustic-crossing frequency during this time is therefore approximately
\begin{align}
\label{eq:dNdtmax}
  \frac{N_{cf}}{\tau_f} & \simeq
 \frac{ 25.4 \, {\rm yr}^{-1} }{\vert \sin{\alpha} \vert^{1/2}} \left( \frac{\theta_*}{\mu{\rm as}} \right)^{-4/3} 
 \left( d\cdot 12 \, {\rm arcsec} \right)^{-1/3}
 \left( \frac{\kappa_*}{\kappa_c} \right)^{1/4}\,\left( {\mathcal{C}_{\star} \over 2.5} \right)^{-1/2}
  \,\left( \frac{v_t}{1000 \, \kms} \right) \, \left( \frac{1.7 \, {\rm Gpc}}{D_S} \right) \,.
\end{align}
For $\kappa_\star = 10^3\, \kappa_c$, the timescale $\tau_f \simeq 3 \, {\rm yr}$, and the maximum microcaustic-crossing frequency is $N_{cf} / \tau_f \simeq 100 \, {\rm yr}^{-1}$.

The widths $s_w$ and $s_f$ are roughly equal for $\kappa_\star \simeq \kappa_c$, but they respectively scale as $\kappa_\star^{1/2}$ and $\kappa_\star^2$ as the surface mass density of microlenses increases. The limit that applies to the expected population of intracluster stars in lensing clusters is $\kappa_\star \gg \kappa_c$, and hence most of the high magnification events do not occur over the narrow width $s_f$ in which the microcaustic-crossing rate is the largest, but over the much broader width $s_w$ with a much lower crossing frequency. \reffig{caustics_network} shows the scales $s_f$ and $s_w$ in this domain for modest values of the surface mass density $\kappa_\star = 17.1 \, \kappa_c = 3.25 \times 10^{-4}$.

We can also use the results in this subsection to investigate whether we can treat each microcaustic crossing as an isolated fold. For a source of radius $R$, this requires that the distance between neighboring caustics exceeds the diameter $2R$. If we impose this condition over the entire band of corrugated microcaustics, where the average angular density of caustics is $N_c/s_w$, the source radius is constrained to $2R/D_S \ll s_w/N_c$, or
\begin{align}
\label{eq:Rsize}
  R & \ll {D_S\over 2}\, \theta_\star\, \kappa_\star^{1/2} \simeq
 1.8 \times 10^4\, R_\odot\, \left( \frac{D_S}{1.7\,{\rm Gpc}} \right)
 \left( \frac{\theta_*}{\mu{\rm as}} \right)
 \left( \frac{\kappa_*}{0.01} \right)^{1/2} ~.
\end{align}
If we require that the source resolves caustics within the narrower band of width $s_f$, where the caustic density is highest, we have a tighter condition:
\begin{align}
\label{eq:Rsizef}
  R & \ll {D_S\over 2}\, \theta_\star\, \kappa_\star^{1/2} \, 
 \left( {\mathcal{C}_{\star} | \sin\alpha | \over 2} \right)^{1/2}
 \left( \kappa_\star \over \kappa_c \right)^{-3/4} ~.
\end{align}
Stars with realistic sizes generally satisfy the first condition, i.e., \refeq{Rsize}, but supergiant stars can violate the second condition, i.e., \refeq{Rsizef}, for the expected intracluster stellar surface mass densities in lensing clusters ($\kappa_\star/\kappa_c \simeq 10^3$, in which case the right-hand side $\simeq 200 \, R_\odot$). Consequently, we expect that the disks of supergiant stars can span, and substantially smooth, multiple peaks in the densest part of the microcaustic network.

\subsubsection{The distribution of magnified micro-images}
\label{sec:magc}

Finally, we discuss the distribution of the large number of micro-images on the image plane for sources within the network of corrugated microcaustics, when the surface mass density of microlenses $\kappa_\star \gg \kappa_c$. This was first worked out by \cite{1986ApJ...306....2K}---we present a condensed version of their argument here. 

Let a source at position $\bt y$ have one of its macro-images at position $\bt x$. Consider a micro-image at a perturbed position $\bt x + \Delta \bt x$ in the presence of microlenses, which has to satisfy the full lens equation,
\begin{align}
\label{eq:deltalenseq}
  \bf y & = \bf x + \Delta \bf x - \bm{\alpha}_{\rm M}(\bf x + \Delta \bt x) - 
  \bm{\alpha}_{\rm ml}(\bfx + \Delta \bt x)\, \mbox{,}
\end{align}
where we have decomposed the deflection $\bm{\alpha} = \bm\nabla \psi$ into a macroscopic contribution $\bm{\alpha}_{\rm M}$ and a stochastic one due to the point masses $\bm{\alpha}_{\rm ml}$. The stochastic term averages to zero because the macroscopic model surface density $\kappa(\bt x)$ includes the mean surface mass density in microlenses, $\kappa_*$, and we include a compensating smooth negative surface density $-\kappa_\star$ in the stochastic term.

If we move the last term on the right-hand side of \refeq{deltalenseq} to the left-hand side, we see that the position $\bt x + \Delta \bt x$ is the image of a source located at $\bt y + \bm{\alpha}_{\rm ml}(\bt x + \Delta \bt x)$ under the {\em macroscopic} lens model. This enables us to derive the magnification-weighted distribution of micro-image positions over all random positions of the microlenses: it is the image, in the macroscopic lens model, of a fictitious source that is centered on $\bt y$ with a surface brightness profile $I(\bt y + \Delta \bt y) \sim p(\bm{\alpha}_{\rm ml} = \Delta \bt y)$ (the distribution of the stochastic microlensing deflection with a characteristic width $s_f$). Note that this is the magnification-weighted image distribution---the unweighted distribution is roughly uniform because each microlens produces at least one image, which can be highly demagnified if it maps to a source position that is much farther from $\bf y$ than $s_f$. This argument, based on \refeq{deltalenseq}, applies even when the Jacobian matrix varies within the region of size $s_f$: when the source is near the macrocaustic, the images consist of a large number of ``arclets'' that are arranged in a highly elongated distribution. The blue dots in the lower panel of \reffig{critical_network} show an example of the micro-images of a source (only images with $\mu > 1$ are shown); they are distributed in an elongated region around the two red dots, which are the macro-images of the same source.

The locations of the micro-images have the maximum spread, $r_f$, when the source is within the width $s_f$ of the region with the maximum density of microcaustics,
\begin{equation}
\label{eq:rf}
 r_f \simeq \left( {2\,s_f\over | d\sin\alpha | } \right)^{1/2} \simeq
 r_w\, \left( {2\, \mathcal{C}_{\star} \over |\sin\alpha | } \right)^{1/2}
       \left( {\kappa_c\over \kappa_\star} \right)^{3/4} ~.
\end{equation}
As discussed earlier, the width $r_w \simeq 0.1 \,{\rm arcsec}$ can be resolvable for characteristic parameter values in lensing clusters. However, since the expected surface mass density of the intracluster stars $\kappa_\star \sim 10^3\, \kappa_c$, the scale $r_f$ is too small to be directly resolved and hence a cloud of micro-images of a magnified star appears as a single point to a telescope. When some of the micro-images significantly brighten (which happens, for example, at microcaustic crossings), the centroid of the unresolved clouds can shift by amounts $\lesssim r_f$. In real observations, diffraction spreads the flux of each unresolved cloud over a characteristic profile. If we have observations of a sufficiently bright source during multiple epochs and the source star crosses microcaustics in between, we can infer centroid shifts from variations in the distribution of fluxes across several pixels.

\subsubsection{Summary: properties of corrugated microcaustics}
\label{sec:sumc}

To conclude, we can summarize the effects of microlenses on the caustic-crossing phenomena in galaxy cluster lenses as follows:
\begin{itemize}

\item In the vicinity of the cluster's macrocritical curve, there is a threshold surface mass density of point masses $\kappa_c$ (in units of $\Sigma_{\rm crit}$) above which the critical curve is strongly disrupted over most of its length. The value of $\kappa_c$ approximately equals the cubic root of the ratio $M_\star/M_C$ of the characteristic microlens mass to the cluster mass projected inside the smooth critical line.

\item The large magnification of the macroscopic lens model enhances the microlensing cross-section, and when $\kappa_\star > \kappa_c$, this causes the macrocaustic to become a network of corrugated microcaustics generated by the microlenses. The width of this network, $s_w$, is proportional to $\kappa_\star^2$. 

\item The characteristic peak magnifications that are achievable during caustic crossings are reduced by the factor $(\kappa_\star/\kappa_c)^{3/4}$ relative to the case of the macroscopic lens model, and are roughly constant within the microcaustic network width $s_w$. When a compact source traverses this width, it crosses a large number of microcaustics $N_c \sim (\kappa_\star/\kappa_c)^{3/2}$. However, the crossing frequency is maximized over a narrower width $s_f$ that is proportional to $\kappa_\star^{1/2}$.

\item The micro-images of a single source have the largest spread on the image plane $r_f \propto \kappa_\star^{1/4}$, when the source is within the region of width $s_f$ with the highest microcaustic density. If there are a number of sources, highly magnified images of any of them are expected to sporadically appear throughout the much larger width $r_w \propto \kappa_\star$ of the network of microcritical curves, which is typically resolvable for lensing clusters.

\end{itemize}

%%%%%%%%%%%%%%%%%%%%%%%%%%%
\section{Numerical Simulations}
\label{sec:numerical}
%%%%%%%%%%%%%%%%%%%%%%%%%%%

In this section, we present the numerical simulation of the light curves of sources that cross a macrocaustic, and demonstrate the effect from a population of microlenses in the vicinity of the macroscopic critical curve.  We adopt the same simplifying assumptions as in \refsec{microlens}: first, we assume that all microlenses have identical masses. Second, we focus on a region on the image plane that is small compared to the cluster lens scale, $\theta_C$, and represent the macroscopic lens model by a fold. Finally, we assume that the microlenses are distributed with a uniform average surface mass density across the region of interest. 

We start with the macroscopic lens model of \cite{Kawamata:2015haa}, remove a surface mass density $\kappa_\star$ from a region of radius $\mathcal{R}$ centered on the critical curve, and redistribute it in point masses (of Einstein radius $\theta_*$ each) with a number density per unit area $n_{\rm m} = \kappa_*/(\pi\,\theta_*^2)$. \reftab{glafic-model} lists the values for the base and the derived parameters of the model, and for the extra parameters used in the simulations. 

Note that the values of $\kappa_\star$ used in our simulations are much lower than the realistic value $\kappa_\star \simeq 0.01$ (see \refapp{application}). The simulations are computationally intensive for very large numbers of microlenses, so our aim is to verify the analytical scalings of \refsec{microlens}, and extrapolate the results to the realistic value $\kappa_\star = 0.01$.

%%%%%%%%%%%%%%%%%%%%
\begin{table}[t]
\begin{center}
\setlength\tabcolsep{9pt}
\begin{tabular}{l|c}
\specialrule{.1em}{.05em}{.05em} 
Parameter & Value(s) \\
\hline
\hline
$\kappa_0$ & 0.83 \\
$s \equiv {\rm Arg}(\hat{\bt s})$ $(^{\circ})$ & -24.80 \\
$\bm\nabla \kappa_{\rm S}$ (${\rm arcmin}^{-1}$) & (-1.42, 3.10) \\
$\bm\nabla \lambda_{\rm S}$ (${\rm arcmin}^{-1}$) & (-0.49, -0.62) \\
$\bt d$ (${\rm arcmin}^{-1}$) & (3.62, -3.41) \\
$\alpha$ $(^{\circ})$ & 71.52 \\
$\theta_*$ (arcsec) & $10^{-6}$ \\
$\kappa_{c}$ & $1.9 \times 10^{-5}$ \\
$\kappa_*$ & $(0.65, 3.25) \times 10^{-4} = (3.4, 17.1) \, \kappa_{\rm c}$ \\
\specialrule{.1em}{.05em}{.05em}
\end{tabular}
\caption{\label{tab:glafic-model} Base and derived parameters of the macroscopic lens model corresponding to the candidate transient in MACS\,J1149.5+2223 and of the microlenses that are used in our simulations. Base parameters are adapted from those of \cite{Kawamata:2015haa}, evaluated at the location of the candidate transient. Gradients are defined in the Equatorial coordinate system of \reffig{magplot}. The unit vector $\hat{\bt s}$ points along the trivial direction of the macroscopic lens map, while the vector $\bt d$ is orthogonal to its critical curve (see \reffig{fold}). $R$ is the source radius, and the quoted value of $\kappa_{\rm c}$ is a convention.}
\end{center}
\end{table}
%%%%%%%%%%%%%%%%%%%%

Due to computational limitations, we include microlenses only within a restricted region on the lens plane near the line of sight to the transient, although in reality they are distributed throughout the lensing cluster. An important question is whether we have adequately sampled the lens plane and included all microlenses that contribute substantially to the total image magnification. To answer this, we recall that highly magnified micro-images are distributed in elongated elliptical regions around the macro-images, which have a maximum size $r_f$ when the source has a separation from the macrocaustic of $\simeq s_f$ (given by \refeq{sf}). From \refeq{rf}, this size is
\begin{align}
  \frac{r_f}{\theta_\star} & \approx
 \left( \frac{2\, s_f}{\theta_\star^2\, d\, \vert \sin{\alpha} \vert } \right)^{1/2}
 = 2.45 \times 10^3 \left( \frac{\kappa_*}{10^{-2}} \right)^{1/4}
 \left( \frac{\theta_\star}{\mu{\rm as}} \right)^{-1/2}
 \left( d\, \vert \sin{\alpha} \vert \cdot 12\, {\rm arcsec} \right)^{-1/2}
 \left( \frac{\mathcal{C}_\star}{2.5} \right)^{1/2} \mbox{.}
\label{eq:pertmacro}
\end{align}
For the largest value of the stellar surface density we simulate $\kappa_\star=3.25\times 10^{-4}$, the ratio $r_f/ \theta_\star \simeq 500$, and as long as we simulate a larger region, we expect to have adequately sampled the image plane in the vicinity of the macrocritical curve. We choose $\mathcal{R} = 1.4 \times 10^{4} \, \theta_\star$ in our simulations.

%%%%%%%%%%%%%%%%%%%%%%%%%%%
\subsection{Algorithm for computing light curves}
\label{sec:lensalg}
%%%%%%%%%%%%%%%%%%%%%%%%%%%

We now describe our algorithm for computing the light curves of moving sources. There are several methods in the literature, the most popular one being inverse ray shooting, coupled with a hierarchical tree code to efficiently compute deflection angles \citep[see, e.g.,][and references within]{Garsden2010181}. This method does not solve the lens equation but gives a map of the magnification at any source position down to the pixel scale. In our case, the typical source-plane scale (e.g., the separation between caustics in \reffig{caustics_network}) is much larger than the source size. Hence, generating high-resolution magnification maps, especially near caustics, would require shooting a prohibitively large number of rays. One alternative is the inverse polygon-mapping technique, which achieves a higher resolution for a given number of rays when compared to the simple ray shooting \citep{2006ApJ...653..942M, 2011ApJ...741...42M}. Instead of those, we choose a different method, outlined below, that explicitly solves the lens equation and finds all images. 

Consider a source in the geometric context of \reffig{fold}, at an initial position $\bt y_0$ on the inside of the macrocaustic. When sufficiently far from the macrocaustic, the two macro-images that dominate the total magnification are not highly perturbed by the microlenses. In addition, there is one micro-image close to each microlens. In this limit, we can start with analytic approximations to the positions of individual images and then refine them using a root-finding procedure to obtain seed image positions for our procedure. 

We now let the source move on a parametric curve $\bt y(\tau)$, starting at $\bt y(\tau = 0) = \bt y_0$, and follow the trajectory $\bt x(\tau)$ of each image. As long as the source curve does not cross a caustic, the image trajectory solves a first-order ordinary differential equation (ODE)
\begin{align}
\label{eq:odeinteg}
  \frac{d}{d\tau} {\bt x}(\tau) & = {\bt A}^{-1}(\bt x(\tau)) \cdot
 \frac{d}{d\tau} {\bt y}(\tau) ~,
\end{align}
where $\bt A$ is the Jacobian of the lens map. The trajectory terminates only when the source crosses a caustic, where the image merges onto a critical curve where $\det {\bt A} = 0$. We numerically evolve the ODE until such a point, taking care to enforce small step sizes when $\det {\bt A} \rightarrow 0$ (which ensures fine resolution near caustics). Assuming the caustic is a fold, the image merges with another one of the opposite parity on the critical curve. We use the limiting solution near a fold \citep{1986ApJ...310..568B} to jump across the critical curve and start a new branch of the image trajectory satisfying \refeq{odeinteg}, but with the $\tau$ integration reversed in direction (i.e., we track the image that merged with the previous one back in time). In this manner, we map out the image trajectory until the source reaches a point that is sufficiently far from the macrocaustic. 
We choose the source-plane curve $\bt y(\tau)$ to be a straight line, and repeat this procedure with all the seed images. Given any source position along the curve $\bt y(\tau)$, we interpolate all of the branches of the image trajectories to read off the micro-images. This algorithm is a variant of the method suggested by \cite{1993MNRAS.261..647L}, who map out trajectories using a sequence of root-finding steps for the lens equation. The equivalence between the methods is a consequence of \refeq{odeinteg} being the Jacobian of the lens equation---using an ODE solver facilitates the adaptive control of the step size. Our method is ultimately slower than inverse ray shooting or polygon mapping for large numbers of stars, but it does not miss any images (by design) and gives point-source light curves of excellent temporal resolution. Our method can also be combined with a hierarchical tree algorithm to speed up the computation of the lensing Jacobian matrix. 

We can also extend the method to obtain light curves for extended sources using the technique described in \cite{1999MNRAS.306..223W}. We save on these additional computations by appealing to the nearly one-dimensional nature of the caustics (see the lower-right panel of \reffig{caustics_network}). This implies that most crossings are simple fold crossings, in which case we can obtain the light curves of extended sources by convolving point-source light curves with a characteristic window function \citep{1987A&A...171...49S, 1991ApJ...379...94M}. This can be inaccurate at peaks due to passages outside cusps---these, however, are rare in the limit of large microlensing optical depth \citep{1992A&A...258..591W}.

%%%%%%%%%%%%%%%%%%%%%%%%%%%%%%%%%%%%%%%%%%%%%%%%%%%%%%%%%%%%%
\subsection{Light curves and caustic-crossing statistics}
\label{sec:lcresults}
%%%%%%%%%%%%%%%%%%%%%%%%%%%%%%%%%%%%%%%%%%%%%%%%%%%%%%%%%%%%%

We now present numerical results for the light curves, and demonstrate and refine the statistical results that we derived in \refsec{microlens}.

%%%%%%%%%%%%%%%%%%%%%%%%%%%%%%%%%%%%%%%%%%%%%%%%%%%%%%%%%%%%%
\begin{figure}[t]
\begin{center}
  \includegraphics[width=0.9\textwidth]{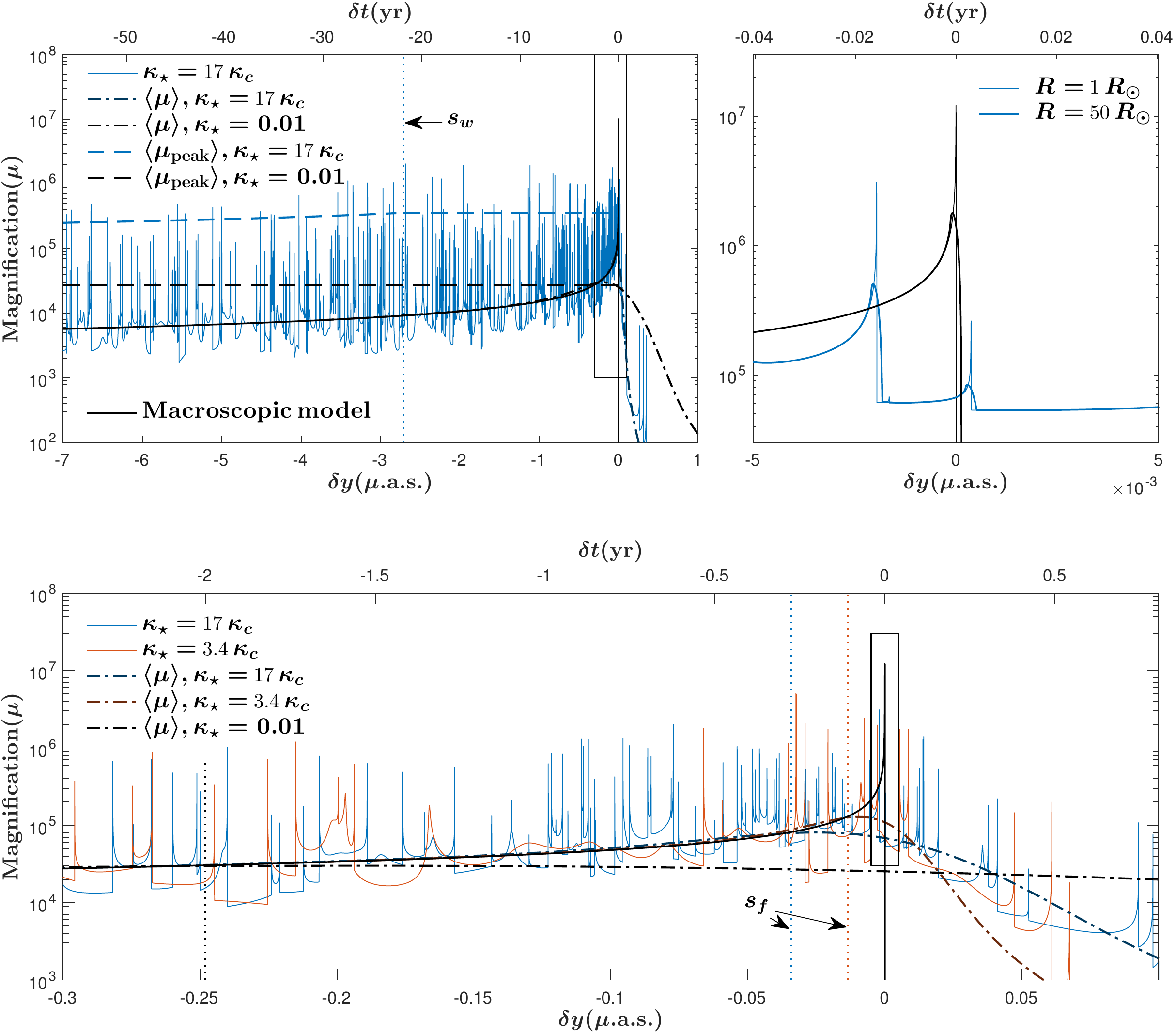}
\caption{\label{fig:lightcurves} Light curves for the lens model of Figures \ref{fig:critical_network} and \ref{fig:caustics_network}: Starting with the top-left figure and moving counterclockwise, panels show details on successively finer scales, with black rectangles marking regions that are zoomed into. Solid black lines are light curves in the macroscopic lens model, for a source radius $R = 1\, R_{\odot}$. The {\color{mlabblue} blue} solid lines are light curves when the microlens surface mass density is $\kappa_\star = 17.1\, \kappa_{\rm c} = 3.25 \times 10^{-4}$, and the {\color{mlabred} red} solid lines in the bottom panel are for $\kappa_* = 3.4\, \kappa_{\rm c} = 6.5 \times 10^{-5}$. The dashed-dotted lines of each color show the mean magnification for random realizations of the microlens positions, as given by \refeq{meanmugen}, and the dashed-dotted black line shows the mean magnification for $\kappa_\star = 0.01$. The dashed curves in the top-left panel are the estimates of the mean peak magnification $\langle \mu_{\rm peak} \rangle$ when $\kappa_\star = 17.1 \, \kappa_{\rm c}$ and $0.01$, respectively, as given by \refeq{microlen_mumax} for a source of radius $R = 1\, R_{\odot}$, and with an extra correction factor of $0.25$ from a fit to the simulations in \reffig{peak_mag}. The characteristic source-plane widths $s_w$ and $s_f$ (as given by Eqs.~\eqref{eq:sw} and \eqref{eq:sf}) mark the beginning of the microcaustic network, and the region of highest caustic-crossing frequency, respectively. The dotted black line in the lower panel marks the source-plane width $s_f$ for $\kappa_\star = 0.01$. The top-right panel, which is zoomed in the most, also indicates the light curve for a source of $R=50\, R_{\odot}$. The timescale on the secondary $y-$axes assumes a velocity $v_t = 1000 \, \kms$ in \refeq{projv}.}
\end{center}
\end{figure}
%%%%%%%%%%%%%%%%%%%%%%%%%%%%%%%%%%%%%%%%%%%%%%%%%%%%%%%%%%%%%

\reffig{lightcurves} presents light curves for particular realizations of microlenses with $\kappa_\star = 17.1\, \kappa_{\rm c} = 3.25 \times 10^{-4}$ (blue lines), and $\kappa_\star = 3.4\, \kappa_{\rm c} = 6.5 \times 10^{-5}$ (red lines) within the lens model of \reffig{magplot}, computed using the method outlined in \refsec{lensalg}. The blue solid lines are light curves for the realization whose critical curves and caustics were shown in Figures \ref{fig:critical_network} and \ref{fig:caustics_network}, respectively. The black solid line is the light curve in the absence of any microlenses. The secondary $y-$axes show the elapsed time if the velocity scale of \refeq{projv} is $v_t = 1000 \, \kms$. 

The dashed-dotted curves show the mean magnification for a given value of $\kappa_\star$, as computed using \refeq{meanmupl} of \refapp{analytics}. The mean magnification is an average over random realizations of microlens positions. In individual realizations, the total magnification fluctuates about this mean both due to microcaustic crossings and Poisson fluctuations in the density of the microlenses. As seen in \reffig{lightcurves},  the magnification is often in valleys that usually dive to values two or three times below the mean, and this is compensated by the microcaustic crossings where the magnification briefly reaches much higher values than the mean.

The top-left panel shows the light curve of a source of radius $R = 1\,R_\odot$ over long timescales. The dotted blue vertical line marks the edge of the band of width $s_w$ (given by \refeq{sw}, and shown in \reffig{caustics_network}), within which microcaustics join to form a network. The typical peak magnification $\mu_{\rm peak}$ at microcaustic crossings increases as the source approaches the network, up to a separation $\sim s_w$, and then remains approximately constant. This agrees with \refeq{microlen_mumax}, which was approximately derived up to a numerical factor of order unity. Later in this section, we infer a value of $0.25$ for this factor by analyzing an ensemble of simulated light curves. The dashed blue and black lines are the result of \refeq{microlen_mumax} for $\kappa_\star = 17\,\kappa_c$ (simulated) and $\kappa_\star = 0.01$ (the realistic value), respectively, after multiplying by an extra factor of $0.25$. We note that for $\kappa_\star = 0.01$, the scale $s_w$ is $\sim 2500\, \mu{\rm as}$, which is well outside the range of the top-left panel.

 The mean magnification for the realistic case with $\kappa_\star=0.01$ (black dashed-dotted line) becomes similar to the typical value of $\mu_{\rm peak}$ (black dashed line) at $| \delta y| \sim 0.3\, \mu{\rm as}$, meaning that when the source is in this region close to the macrocaustic, the magnification of the two bright images of a microcaustic crossing is about equal to the sum of the magnifications of the numerous other micro-images.
 
The lower panel of \reffig{lightcurves} zooms into the black rectangle that is marked in the top-left panel. The values of $\simeq s_f$, as given by \refeq{sf}, are indicated by the blue and red dotted vertical lines for the two simulations. The black dotted vertical line shows the value of $s_f = 0.25 \, \mu{\rm as}$ for the realistic case. The blue and red dashed-dotted lines are the mean magnification for the two values of $\kappa_\star$ of the light curves, and the dashed-dotted black line is the mean magnification for $\kappa_\star = 0.01$. The frequency of microcaustics is maximum near the macrocaustic, within a band of width $\sim s_f$, and drops outside this region. The visual impression from \reffig{lightcurves} is that the width of the region of roughly constant microcaustic density is a few times $s_f$, which will be confirmed below.

Our expectation is that sources with typical sizes of a few solar radii are not large enough to cross multiple microcaustics at once (see \refeq{Rsize}). Hence, the main effect of the source size will be to regulate the peak magnification $\mu_{\rm peak}$ at individual crossings. This is illustrated in the top-right panel of \reffig{lightcurves}, which zooms into the black rectangle that is marked in the bottom panel, and shows the macro and micro light curves for two very different source radii. As discussed in \refeq{Rsize}, typically stars cross each microcaustic separately, except for giant stars in the region of highest microcaustic frequency for $\kappa_\star\sim 0.01$. We note that clustering of microcaustics, caused by Poisson fluctuations in the density of microlenses, may increase the probability of overlapping microcaustic crossings; we neglect this here and present numerical results treating microcaustics as a one-point statistic.

Our analytical estimates in \refsec{caustics} lead us to expect that the microcaustic-crossing frequency scales as $(\kappa_\star/\kappa_c)^{3/4}\, \mathcal{C}_\star^{1/2}$ when expressed in terms of the ratio $\delta y/s_f$, where $\delta y$ is the separation from the macrocaustic on the source plane, and $s_f$ is the width of highest microcaustic density of \refeq{sf}. Taking out this dependence, this crossing frequency should have a unique functional form. We verify this using an ensemble of numerical simulations containing $300$ and $150$ realizations of microlenses to accumulate enough statistics, with the same values of the surface mass density as in \reffig{lightcurves}, $\kappa_\star = 3.4 \, \kappa_{\rm c}$ and $17.1 \, \kappa_{\rm c}$, respectively.
The algorithm we use to solve the lens equation naturally gives the crossing locations, which is where we reverse the direction of the time integration in \refeq{odeinteg}.

%%%%%%%%%%%%%%%%%%%%%%%%%%%%%%%%%%%%%%%%%%%%%%%%%%%%%%
\begin{figure}[t]
\begin{center}
  \includegraphics[width=12cm]{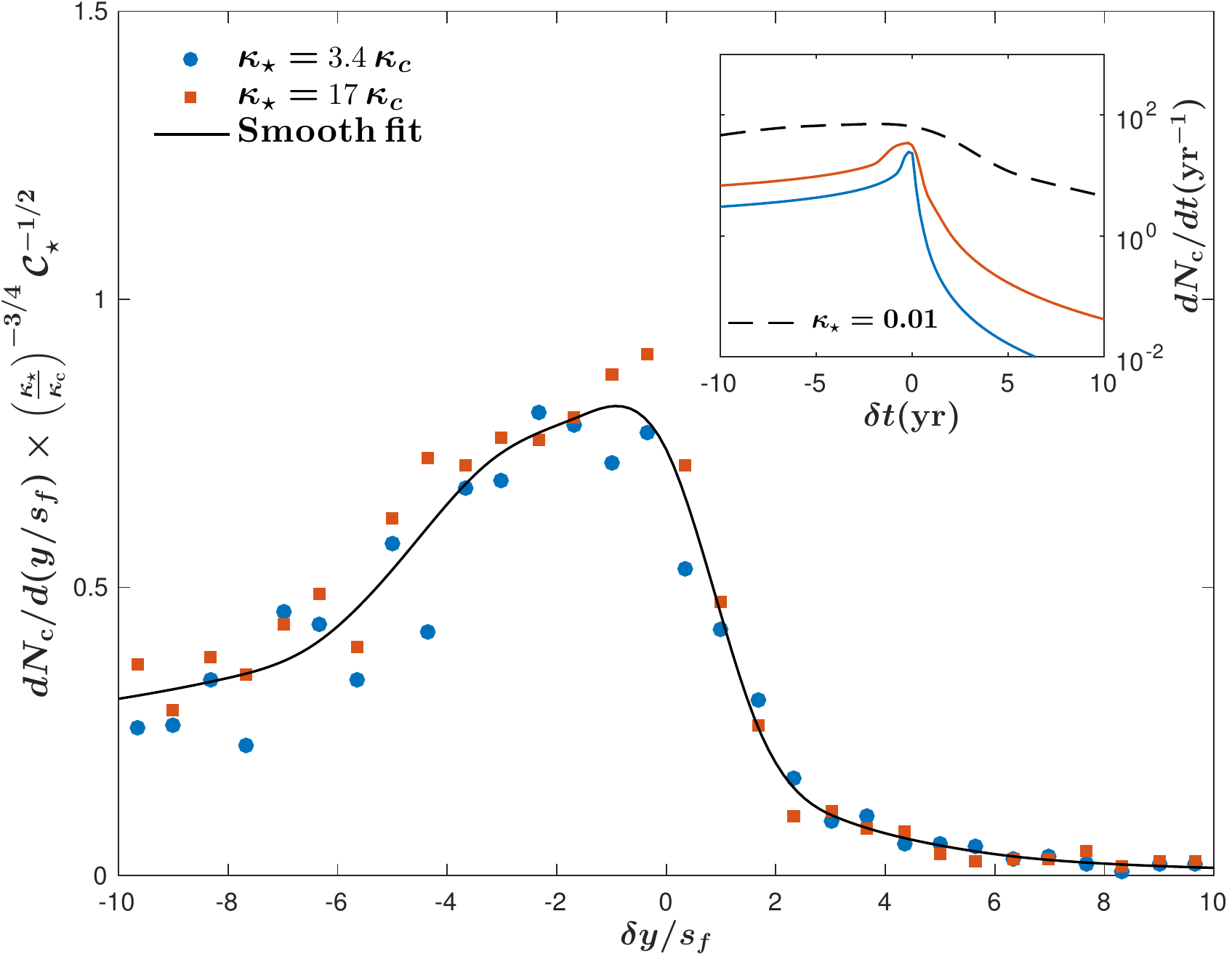}
\caption{\label{fig:ccross_freq} Frequency of microcaustic crossings (with no magnification threshold) as a function of the source displacement from the macrocaustic divided by $s_f$ (from \refeq{sf}). The {\color{mlabblue} blue} circles and {\color{mlabred} red} squares are the values measured from an ensemble of light curves for microlens surface mass densities of $\kappa_\star = 3.4\, \kappa_{\rm c}$ and $17.1\, \kappa_{\rm c}$, respectively. Crossing frequencies have been multiplied by the factor needed to scale them to a common form, based on the analytical estimate in \refeq{dNdtmax}. The solid black line is a smoothing cubic spline fit to the numerical data. The inset shows crossing frequencies in physical units for the two simulated values of $\kappa_\star$, and for its expected value in the intracluster region, $\kappa_\star = 0.01$, when the velocity scale of \refeq{vrel} is $v_t = 1000 \, \kms$.}
\end{center}
\end{figure}
%%%%%%%%%%%%%%%%%%%%%%%%%%%%%%%%%%%%%%%%%%%%%%%%%%%%%%

\reffig{ccross_freq} shows the scaled crossing frequencies of all microcaustics (i.e., no magnification threshold is imposed) as a function of the scaled displacement from the macrocaustic. The blue circles and red squares indicate the results for the two values of $\kappa_\star$ used in our simulated light curves. There is significant statistical scatter in the estimated frequencies, but they are distributed about a common curve. Far from the crossing, on the inside of the caustic, the frequencies follow the macroscopic magnification curve, i.e., $dN/dt \sim |t|^{-1/2}$. They peak at a separation $\sim s_f$, and rapidly decline outside the macrocaustic, falling as $dN/dt \sim |t|^{-2}$.

The black solid line in \reffig{ccross_freq} is a smoothing cubic spline curve that captures this behavior. The inset shows the crossing frequencies in physical units for the two cases we simulated. As in \reffig{lightcurves}, we assume that the velocity scale $v_t = 1000 \, \kms$. The dashed black line is an extrapolation to the case $\kappa_\star = 0.01$, obtained by the expected rescaling of the solid black line.

 The scaled peak crossing frequency predicted by our analytical equation (\ref{eq:ncf}) is $N_{cf}\, (\kappa_\star/\kappa_c)^{-3/4}\, \mathcal{C}_\star^{-1/2} = (2/|\sin\alpha |)^{1/2} = 1.45$ (for the model parameters in \reftab{glafic-model}). In comparison, the maximum scaled frequency in \reffig{ccross_freq} is $\sim 0.8$. We note, however, that the width of the region with roughly constant microcaustic frequency is $\sim 4\,s_f$ (confirming the visual impression from the example light curves in \reffig{lightcurves}), and the predicted mean scaled frequency in this region is $1.45/2$, in very good agreement with the numerical result. We therefore conclude that the analytic predictions from \refsec{freqc} are in excellent agreement with our simulations if we replace the width of the region of a roughly constant microcaustic density by $4\,s_f$.

%%%%%%%%%%%%%%%%%%%%%%%%%%%%%%%%%%%%%%%%%%%%%%%%%%%%%%%%%%%%%
\subsection{Microcaustic peak magnification distribution}
\label{sec:peakmagdist}
%%%%%%%%%%%%%%%%%%%%%%%%%%%%%%%%%%%%%%%%%%%%%%%%%%%%%%%%%%%%%

\begin{figure}[t]
\begin{center}
  \includegraphics[width=12cm]{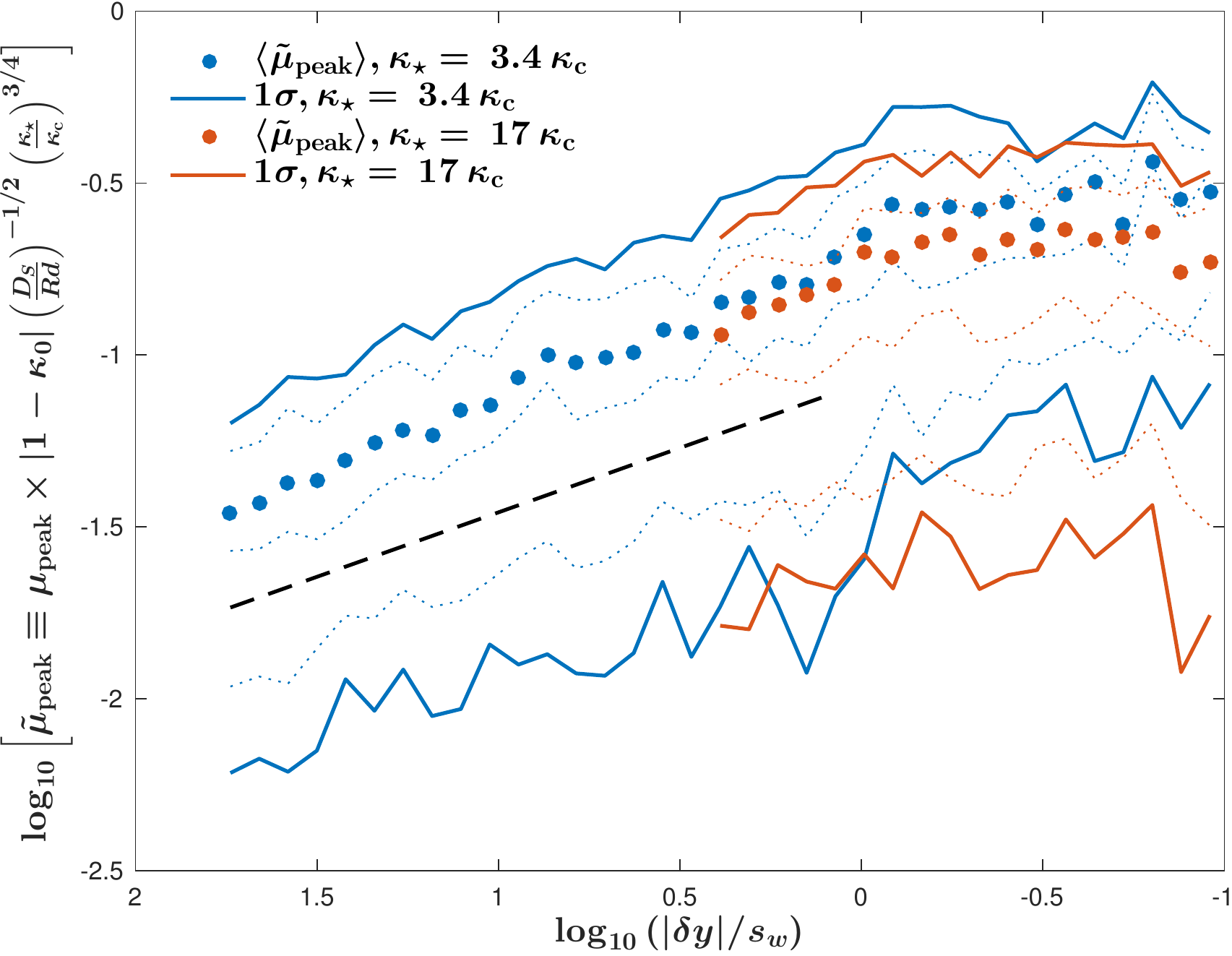}
\caption{\label{fig:peak_mag} Probability distribution of the peak magnification at microcaustic crossings as a function of the scaled source displacement from the macrocaustic, measured from the same ensemble of simulated light curves as in \reffig{ccross_freq}. The horizontal axis has been rescaled in units of the source-plane width $s_w$ from \refeq{sw}, while the vertical axis shows the scaled peak magnification $\tilde{\mu}_{\rm peak} \equiv \mu_{\rm peak} \vert 1 - \kappa_0 \vert \left( D_S/R d \right)^{-1/2} \left( \kappa_\star/\kappa_{\rm c} \right)^{3/4}$ (according to \refeq{microlen_mumax}, this definition should scale out the dependence on $\kappa_\star$). Dots mark the mean value of the scaled peak magnification, which levels off at a value $\simeq 0.25$ within the band $s_w$. The upper and lower solid lines mark the $16^{\rm th}$ and $84^{\rm th}$ percentiles, respectively. In addition, thin dotted lines mark the $25^{\rm th}$, $50^{\rm th}$, and $50^{\rm th}$ percentiles. We plot two cases $\kappa_\star = 3.4\,\kappa_c$ (blue) and $\kappa_\star = 17\,\kappa_c$ (red). \refeq{microlen_mumax} predicts that the mean follows a power-law behavior with displacement outside $s_w$, and the predicted exponent of $3/8$ agrees with the trend as shown by the dashed black line.}
\end{center}
\end{figure}

Next, we investigate the peak magnification of microcaustic crossings. \reffig{peak_mag} shows the statistics for $\mu_{\rm peak}$ as a function of displacement from the macrocaustic in our ensemble of simulated light curves, for microlens surface mass densities of $\kappa_\star = 3.4 \, \kappa_{\rm c}$ (black) and $17.1 \, \kappa_{\rm c}$ (red). The figure includes only peaks on the inside of the macrocaustic. The displacement $\delta y$ is rescaled to the thickness $s_w$ of the microcaustic network, and magnifications are rescaled according to \refeq{microlen_mumax}. For each bin along the horizontal axis, the bins along the vertical direction are color-coded to represent a histogram of the probability distribution of $\mu_{\rm peak}$ within that displacement range. The figure also shows the mean values of $\mu_{\rm peak}$ and the one-sigma limits of the distribution when all microcaustics are weighted equally. 

%%%%%%%%%%%%%%%%%%%%%%%%%%%%%%%%%%%%%%%%%%%%%%%%%%%%%%
%%%%%%%%%%%%%%%%%%%%%%%%%%%%%%%%%%%%%%%%%%%%%%%%%%%%%%

 First, we observe that, as predicted by our analysis in \refsec{microlens}, the typical peak magnification rises as the source approaches the caustic network and levels off within $s_w$. The behavior of the mean in \reffig{peak_mag} is in excellent agreement with the power law in \refeq{microlen_mumax}. Second, the similarity of the distributions of the scaled $\mu_{\rm peak}$ for the two surface mass densities demonstrates the validity of the scaling with $(\kappa_\star/\kappa_{\rm c})^{3/4}$. Thirdly, we see that the true mean value $\langle \mu_{\rm peak} \rangle$ is lower than the analytical estimate of \refeq{microlen_mumax} (equal to one after the rescaling in this plot), approximately by a factor of four, and microcaustics with a peak magnification as high as that predicted by \refeq{microlen_mumax} are only a few percent of the total. We infer a mean reduction factor of $\simeq 0.25$ as compared to the analytical estimates, from the mean of the red and blue circles within a width $s_w$ in \reffig{peak_mag}. The second and third points are crucial for extrapolating to the realistic surface mass density $\kappa_\star = 0.01$ (dashed black line in the top panel of \reffig{lightcurves}).

Overall, our analytical estimates from \refsec{microlens} are in good agreement with the numerical results after we increase the width of constant microcaustic density to $4\,s_f$ and reduce the mean peak magnification of caustic crossings to $0.25\, \mu_{\rm peak}$ from \refeq{microlen_mumax}. This allows us to make quantitative predictions for the light-curve statistics for the case $\kappa_\star = 0.01$, which we have not been able to simulate directly. According to our estimates in \refapp{application}, this is the expected surface mass density of intracluster stars in the case of the candidate event in MACS\,J1149.5+2223.

For this value of $\kappa_\star$, the network of corrugated microcaustics is remarkably wide, $s_w \simeq 0.0025\,{\rm arcsec}$. As a source approaches the macrocaustic, the mean magnification $\langle \mu \rangle$ increases from $\simeq 300$ at the onset of the band all the way to $\simeq 1.5\times 10^4$ at the inner band $4\,s_f \simeq 1\,\mu{\rm as}$. During this process, we anticipate that the frequency of microcaustic crossings increases from $\simeq 1.4\,{\rm yr}^{-1}\,(v_t/1000\,\kms)$ to $\simeq 70\,{\rm yr}^{-1}\,(v_t/1000\,\kms)$. Throughout this region, the average peak magnification at individual microcaustic crossings is constant at $\langle \mu_{\rm peak} \rangle \simeq 9\times 10^3\,(10 R_\odot/R)^{1/2}$, with a few percent of these crossings reaching values as high as $\mu_{\rm peak}\sim 4\times 10^4\, (10 R_{\odot}/R)^{1/2}$. A source moving at $v_t=1000 \kms$ would take $2\times 10^4$ years to cross the width $s_w$, and only eight years to cross the width $4 \, s_f$, with each crossing event lasting about four hours for $R=10 \, R_\odot$.

When $\langle \mu_{\rm peak} \rangle$ substantially exceeds $\langle \mu \rangle$, the two merging images dominate the total observed flux at each microcaustic crossing, and produce high peaks that would stand out in high-cadence observations. Conversely, if $\langle \mu_{\rm peak} \rangle \lesssim \langle \mu \rangle $, the flux in the two merging images can be comparable, or even subdominant, to the sum of all the other images, in which case the microcaustic crossings would cause minor peaks on a continuously varying image of a microlensed star. For $\kappa_\star = 0.01$ and $R=10\,R_{\odot}$, the latter case occurs only for sources at a separation $|\delta y| \sim s_f \ll s_w$, which will therefore be rare compared to sources with $\langle \mu_{\rm peak}\rangle \gg \langle \mu \rangle$.

%%%%%%%%%%%%%%%%%%%%%%%%%%%
\section{Constraints on MACHOs and small-scale mass inhomogeneities}
\label{sec:machos}
%%%%%%%%%%%%%%%%%%%%%%%%%%%

 We have shown above that the corrugated microcaustics, which cause extreme magnification microlensing events of background stars in galaxy clusters, sensitively depend on the population of microlenses that account for a small fraction of the total lensing mass. We now apply the analytic results of \refsec{analyt} to discuss how the effect of a population of MAssive Compact Halo Objects (MACHOs), apart from the known intracluster stars, on these microlensing events may be used to extend the existing limits on the mass fraction of MACHOs. For simplicity, we assume that all MACHOs have the same mass $M_m$, with a corresponding Einstein radius $\theta_m$ from \refeq{einsteinpt}. Throughout, we assume the physical parameters of the candidate event in MACS\,J1149.5+2223.
 
\begin{figure}[t]
  \begin{center}
    \includegraphics[scale=0.7]{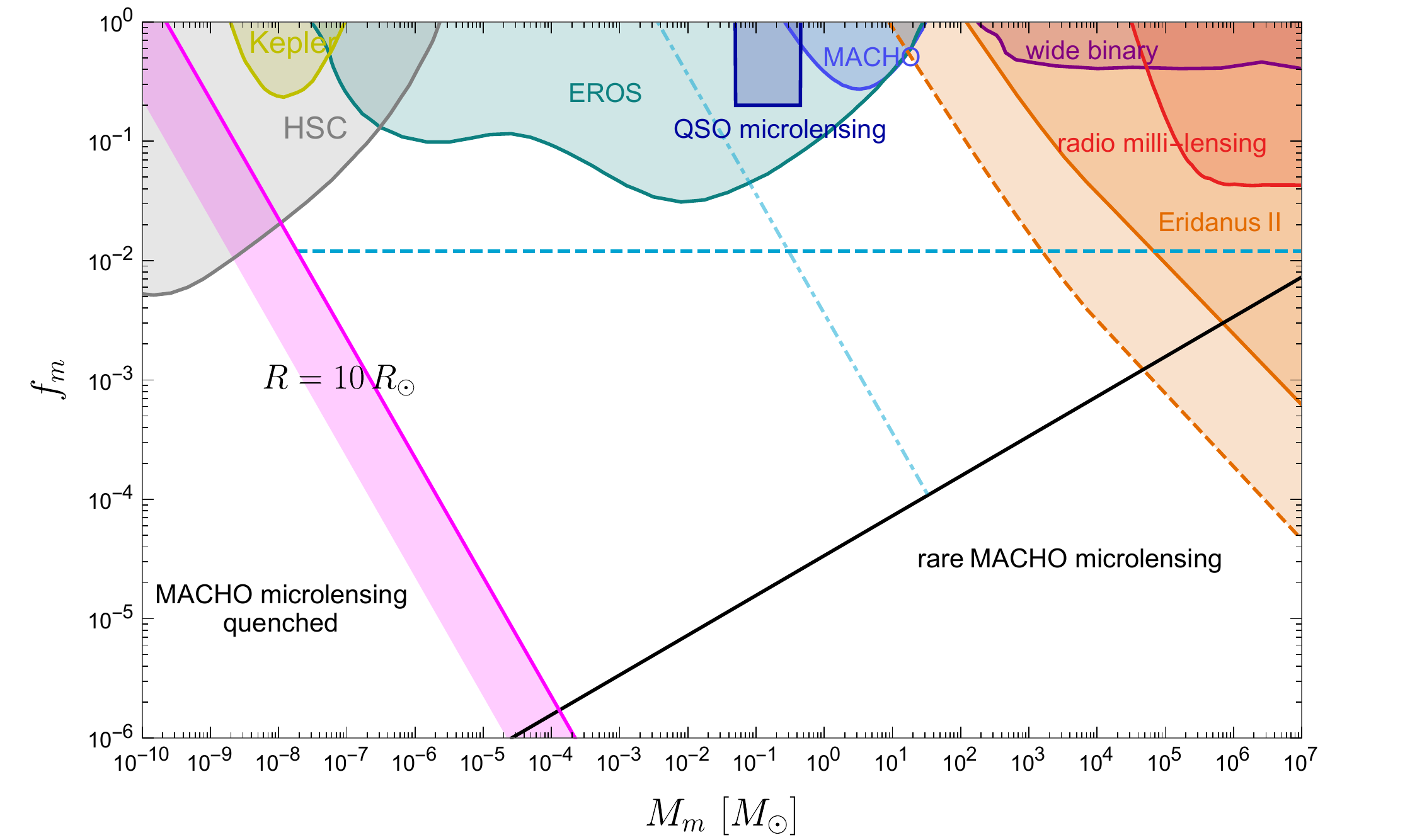}
    \caption{\label{fig:macholimit}
Mass fraction in MACHOs $f_m$ versus the MACHO mass $M_m$.
Black solid line: minimum abundance of MACHOs required to create their own corrugated caustics. Cyan dashed line: minimum abundance of MACHOs that dominate the source-plane width $s_w$ of the microlensing caustic network. Cyan dashed-dotted line: minimum abundance of MACHOs that dominate the source-plane width $s_f$ of the band with the highest caustic density. Purple solid line and shaded band: minimum abundance of MACHOs whose effects are resolved by a source size $R=10\, R_{\odot}$; up to a small numerical factor, this is also close to the minimum abundance of MACHOs for their microlensing deflections to broaden the stellar caustic-crossing peaks beyond the effect of the finite source size (for $\mathcal{C}^2_m = 10$). We use fiducial parameter values for the candidate event in MACS\,J1149.5+2223, with a stellar mass $M_\star = 0.3\,M_\odot$ and $\kappa_\star = 0.01$. We also overplot existing observational constraints: microlensing surveys from MACHO \citep{2001ApJ...550L.169A}, EROS \citep{2007A&A...469..387T}, {\em Kepler} \citep{PhysRevLett.111.181302}, and HSC \citep{2017arXiv170102151N}, quasar microlensing \citep{Mediavilla:2017bok}, Galactic wide binaries \citep{2009MNRAS.396L..11Q}, millilensing of compact radio sources \citep{Wilkinson:2001vv}, and dynamical heating of a star cluster in the dwarf galaxy Eridanus II \citep[][solid line for conservative consideration and dashed line for a more stringent limit]{2016ApJ...824L..31B}.}
  \end{center}
\end{figure}

 \reffig{macholimit} shows the relevant regions in the parameter space of the fraction of the mass fraction in the dark matter of MACHOs, $f_m$, versus their mass $M_m$. The filled regions in the upper part of the plot show the present constraints on the mass fraction $f_m$ from microlensing events in the EROS \citep{2007A&A...469..387T}, MACHO \citep{2001ApJ...550L.169A}, {\em Kepler} \citep{PhysRevLett.111.181302}, and HSC \citep{2017arXiv170102151N} surveys; the existence of very wide binary stars \citep{2009MNRAS.396L..11Q}; the presence of a star cluster in the Eridanus II dwarf galaxy \citep{2016ApJ...824L..31B}; quasar microlensing \citep{Mediavilla:2017bok}; and millilensing of compact radio sources \citep{Wilkinson:2001vv}.

%%%%%%%%%%%%%%%%%%%%
%%%%%%%%%%%%%%%%%%%%

The black, cyan, and purple lines in \reffig{macholimit} mark the transition regions where the effect of the MACHOs on the network of corrugated microcaustics changes, according to our analytical estimates, in the following manner:

\begin{enumerate}

\item The macrocaustic of the cluster lens is affected along most of its
length when the contribution to the convergence from microlenses is
above the threshold value, i.e., when $\kappa_m = f_m\,\kappa_0 > \kappa_c$
(see \refeq{kappac}). Ignoring first the presence of intracluster stars,
this implies
\ba
f_m \gtrsim \frac{1.4 \times 10^{-4}}{\kappa_0}\,
\left( \frac{M_m}{100\,M_\odot} \right)^{1/3}\,
\left( 12\,{\rm arcsec}\cdot d \right)^{2/3}\,
\left( \frac{2.55\,{\rm Gpc}}{D_{\rm eff}} \right)^{1/3} ~,
\ea
where $D_{\rm eff} = D_L\,D_S/D_{LS}$. This is shown by the black solid
line. The presence of MACHOs alters the original macrocaustic of the
cluster everywhere above this line, and would therefore be detectable
in typical caustic-crossing light curves in the absence of
intracluster stars. In the presence of microlensing by intracluster
stars, the two types of microcaustics at different angular scales
become superposed, and we need other criteria to clearly distinguish the effects of the MACHOs from those of intracluster stars, although if $\kappa_m < \kappa_\star$ this is possible only when their masses are sufficiently different.

\item When the abundance of MACHOs is larger, they dominate the total width $r_w$ of the network of corrugated critical curves (\refeq{rw}) compared to intracluster stars. This happens when
\ba
f_m \gtrsim \frac{0.01}{\kappa_0}\,\left( \frac{\kappa_\star}{0.01} \right) ~ ,
\ea
as marked by the cyan dashed line, which is drawn assuming $\kappa_\star=0.01$ and $\kappa_0 = 0.8$.
Note that the width $r_w$ depends only on the surface density
in the microlenses and not on their mass, and can be measured from the distribution of highly magnified images in several caustic-crossing events on the image plane, because it is already resolvable for
the expected surface density in intracluster stars. Above the cyan
line, the width $r_w$ is increased compared to the expected value from
intracluster stars; the mass of the MACHOs can then be inferred from the
characteristic magnifications and frequencies of the microlensing
events.

\item The frequency of caustic crossings peaks in the inner band of width $s_f$ (\refeq{sf}) on the source plane. The width $s_f$ is dominated by microlensing deflections from the MACHOs instead of the stars if
\ba
f_m \gtrsim \frac{10^{-4}}{\kappa_0}\,
\left( \frac{\kappa_\star}{0.01} \right)\,
\left( \frac{100\,M_\odot}{M_m} \right)\,
\left( \frac{M_\star}{M_\odot} \right) ~,
\ea
as marked by the cyan dashed-dotted line. In general, for very large MACHO masses, the peaks due to stellar microcaustics can be viewed as perturbations on top of the MACHO-induced peaks. We also need to take into account that the ordinary population of intracluster stars contains stellar black holes made in the core collapse of massive stars, so any additional population of MACHOs would have to be distinguished from these expected stellar remnants.

\item MACHOs whose masses are much lower than typical stellar masses can generate corrugated microcaustics on smaller angular scales. Thus, their presence would lead to more frequent peaks of lower magnification. However, for low values of the mass $M_m$, the smaller-scale microcaustics are smoothed out by the finite source radius $R$. As derived in \refeq{Rsize} applied to MACHOs, the condition to avoid this smoothing over the entire width $s_w$ is $2R/D_S < \theta_m (f_m \,\kappa_0)^{1/2}$, or
\ba
\label{eq:cond4}
 f_m \gtrsim \frac{2\times 10^{-4}}{\kappa_0}\,
 \left( \frac{10^{-6}\,M_\odot}{M_m} \right)\, 
 \left( \frac{D_{\rm eff}}{2.55\,{\rm Gpc}} \right)\,
\left( \frac{1.7\,{\rm Gpc}}{D_S} \right)^2\,
\left( \frac{R}{10\,R_\odot} \right)^2 ~.
\ea
This condition is shown by the solid purple line in
\reffig{macholimit},
for the fiducial values in our equation. To the left of this line, the microcaustic crossings due to MACHOs are increasingly difficult to observe as they are smoothed by the source size. The light curves then track the average magnification over all possible MACHO positions, keeping fixed the specific realization of the intracluster star positions. The caustic network becomes gradually more resolved as we move to the right of the line: as the MACHO mass $M_m$ is increased at fixed $f_m$, the density of the caustics decreases and more of the network is resolved at fixed $R$, until the width $s_f$ is resolved as well.

There is a second, lower, threshold due to the competition between the smoothing of the stellar microcaustics due to the deflections by MACHOs (over an angular scale $s_f$ applied to MACHOs) and the smoothing due to the source size $R$. Both effects are equally important when $2R/D_S < \mathcal{C}_m \theta_m (\kappa_0\, f_m)^{1/2}$ (where $\mathcal{C}_m$ is the analogous quantity for MACHOs to $\mathcal{C}_\star$ for the intracluster stars, as given in \refeq{sf}). This results in the condition
\ba
\label{eq:cond4b}
 f_m \gtrsim \frac{2\times 10^{-3}}{\kappa_0 \, \mathcal{C}_m^2}\,
\left( \frac{10^{-7}\,M_\odot}{M_m} \right)\,
\left( \frac{D_{\rm eff}}{2.55\,{\rm Gpc}} \right)\,
\left( \frac{1.7\,{\rm Gpc}}{D_S} \right)^2\,
\left( \frac{R}{10\,R_\odot} \right)^2 ~.
\ea
This is nearly identical to \refeq{cond4}, except for the
factor $\mathcal{C}_m^2$ in the denominator. A typical value is $\mathcal{C}_m^2 \simeq 10$
(the number of MACHOs contributing to the deflection variation $s_f$ in
a given stellar caustic is
$\sim f_m\,(\kappa_0/\kappa_\star)\, (M_m/M_\star)^{1/2}$). Given the uncertainties in the numerical pre-factors of our equations, we use a shaded band in \reffig{macholimit} to mark a smooth transition region between the conditions in Eqs.~\eqref{eq:cond4} and \eqref{eq:cond4b}.
\end{enumerate}

  In summary, there is a very large region of the $f_m - M_m$
parameter space in \reffig{macholimit} in which MACHOs can be ruled out or
detected by observations of caustic-crossing stars in lensing
clusters, and which has not so far been constrained by other observations.
First of all, the presence of MACHOs with a contribution to the
convergence much above $\kappa_\star$ would be made obvious because the
most highly magnified images during microcaustic crossings would be spread
over a wide band next to the critical curve of the macroscopic lens model. The characteristic angular extent of this spread should be easily resolvable by telescopes such as {\em HST} and {\em JWST} if the total convergence of microlenses is greater than about $2 \times 10^{-3}$. This characteristic scale could be inferred from the distribution of the locations of detected transients, which can then constrain MACHOs contributing a convergence much greater than that of the stars. A more careful analysis that takes into account the finite cadence of magnitude-limited observations would allow for much more accurate constraints.

The presence of MACHOs can be discerned even in the region $f_m\, \kappa_0 < \kappa_\star$, when their masses are sufficiently different from those of intracluster stars. MACHOs with low masses further corrugate the microcaustics of the stellar population and yield more numerous caustic crossings of lower magnification. In addition, low-mass MACHOs can smooth the stellar microcaustics due to their additional microlensing deflection, even when their individual microcaustics are not resolved by the source sizes. The shaded purple area marks the minimum detectable MACHO masses using this method. Accurate limits from observational data would require modeling of the source star.

If a substantial fraction of the dark matter is in massive MACHOs, the increased microlensing deflections increase the narrow width $s_f$ of the band with the highest caustic-crossing density. This effect is important to the right of the dashed-dotted cyan line and affects the structure of most of the macrocaustic as long as we are above the solid black line. There are therefore ample opportunities for exploring the presence of MACHOs over a broad mass range and down to very low dark matter fractions by carefully observing the distribution and light curves of highly magnified, caustic-crossing images of lensed stars.

Finally, we point out that the high sensitivity of the corrugated caustic structures to small fractions of compact dark matter also extends to other small-scale fluctuations in the surface density. As an example, we consider the possibility that part of the dark matter is made of ultralight axions, which would propagate in dark matter halos as classical scalar waves~\citep{PhysRevLett.85.1158, Schive:2014dra, Hui:2016ltb}. An axion mass $m_a\sim 10^{-22}\, {\rm eV}$ has been hypothesized to possibly explain some discrepancies of the distribution of dark matter in dwarf galaxies in comparison to simple predictions from N body simulations for cold dark matter \citep{2016MNRAS.460.4397C, 2016JCAP...07..048U}. This axion mass is significantly constrained by the latest measurements of the Lyman-alpha forest power spectrum \citep{2017arXiv170304683I}, although these constraints are not severe if the dark matter is not entirely composed of ultralight axions. 
This mass corresponds to a de Broglie wavelength of
the axion $\lambda_a = \hbar/(m_a\, v) \sim 20\, {\rm pc}$, for a typical cluster orbital velocity $v\sim 1000 \kms$. The axion dark matter in a cluster would therefore consist of scalar waves with a density that fluctuates over this scale $\lambda_a$. At a projected radius $r\simeq 50\, {\rm kpc}$, the incoherent superposition of density fluctuations leads to fluctuations in the convergence along the line of
sight of typical size
\begin{equation}
 \Delta\kappa_a \simeq \kappa_0\, (r/\lambda_a)^{-1/2}\,f_a \simeq 0.01\,f_a ~,
\end{equation}
where $f_a$ is the fraction of mass in the form of ultralight axions. These fluctuations in $\kappa$ would be moving transversely at a characteristic velocity $v$, implying variations on a timescale of $\sim 2\times 10^4$ years. In analogy to the effect of MACHOs, the network of corrugated microcritical curves would be displaced typically by an angle $\Delta\kappa_a/d \sim 0.2\,f_a\, {\rm arcsec}$, which can be resolved and would vary randomly along the critical curve of the cluster. These random fluctuations in the projected surface density would also cause modulations in the microcaustic densities on the source plane.

This suggests that detailed observations of caustic-crossing events provide a unique opportunity for discovering small perturbations, not only due to MACHOs, but other exotic possibilities for dark matter components such as ultralight axions, in the surface density of lensing clusters. Ultimately, we need detailed studies that carefully account for observational biases in order to distinguish these effects from those of ordinary cluster galaxies and cold dark matter satellites.

%%%%%%%%%%%%%%%%%%%%%%%%%%%
\section{Conclusions and Discussion}
\label{sec:concl}
%%%%%%%%%%%%%%%%%%%%%%%%%%%

Caustic-crossing stars reaching extreme magnifications in cluster-lensing systems were predicted long ago as a new type of gravitational lensing phenomenon at cosmological distances. Recently, observational evidence for these transients has surfaced in imaging surveys of strong-lensing
clusters with HST. We have shown in this paper that intracluster stars in the foreground play a crucial role in this phenomenon owing to a dramatic enhancement of the microlensing cross-section near macrocritical lines, where the Jacobian matrix of the macrolensing is nearly degenerate.

We find that the expected surface density of intracluster stars (formed due to tidal stripping of cluster galaxies) typically gives a contribution to the convergence of $\kappa_\star \sim 0.01$ at $\sim 50\,{\rm kpc}$ from the center of a lensing cluster. This surface density far exceeds the threshold value $\kappa_c$ necessary to strongly perturb the macrocaustic everywhere and convert it into a band of corrugated microcaustics. For typical parameters of lensing clusters, the characteristic width $s_w$ of this band is a few milliarcseconds, and a source crosses it over a duration of $\sim 10^4$ years. In the period, the source crosses a large number of microcaustics $N_c \sim 6 \times 10^4$, with typical
values of the maximum magnification $\mu_{\rm peak}\sim 10^4\, (R/10 R_\odot)^{-1/2}$, and peak durations of $\sim 5\, {\rm hr}\, (R/10 R_\odot)$. The distribution of peak magnifications during the microcaustic crossings remains roughly constant along the entire width $s_w$. However, the caustic-crossing frequency is not constant, with an average value of $\sim 2\, {\rm yr}^{-1}$ and a maximum of $70\, {\rm yr}^{-1}$ that is achieved over a narrow width $4\,s_f \sim 1\, \mu{\rm as}$, which is crossed over $\sim 10$ years. Because $\mu_{\rm peak}$ is essentially constant over the full width $s_w$, the vast majority of the observed highly magnified images of microlensed stars should be in a region with low microcaustic density, where the caustic-crossing rates are $\sim 1\, {\rm yr}^{-1}$.

The fact that stellar microlensing reduces the peak magnifications achieved implies that only intrinsically luminous stars can be seen by present telescopes during caustic-crossing events. For example, an AB-magnitude limit of $27$ with $\mu_{\rm peak} = 10^4$ implies a source magnitude of $37$ prior to magnification, which at redshift $z_S \simeq 1.5$ implies a luminosity $L\sim 10^5\, L_{\odot}$---the radius cannot be much larger to avoid a further decrease in $\mu_{\rm peak}$. Massive main-sequence stars and blue giant stars are the best sources for caustic-crossing events, and thus the number of detected microcaustic crossings should rapidly increase as new observations reach fainter limiting magnitudes. Microlensing by intracluster stars dramatically increases the rate of caustic crossings and also the area where they occur on the image plane.

Microlensing by intracluster stars breaks the two highly magnified macro-images of the source into ``clouds'' of micro-images, each of which has a maximum angular extent of $r_f \sim 0.003\, {\rm arcsec}$ that is reached when the source is at a distance of $s_f \sim 0.25 \, \mu{\rm as}$ from the macrocaustic. Whenever the source encounters a microcaustic during its motion relative to the lens, micro-images with maximum magnifications $\sim \mu_{\rm peak}$ appear within the clouds, which leads to a stochastic `jitter' of amplitude $\lesssim r_f$ in the image centroids. As the source crosses a width $s_w \sim 2500 \, \mu{\rm as}$, in addition to exhibiting this intermittent `jitter,' the clouds systematically drift a distance of $r_w \sim 0.2$ arcsec on the image plane. The typical peak magnification $\mu_{\rm peak}$ is constant within, and decreases beyond, this separation. The scale $r_f$ is usually not resolvable without multi-epoch observations, but the width $r_w$ can be resolved and measured if several microcaustic-crossing events by different stars in a source galaxy are observed (which is likely to occur given the large width $s_w$ of the corrugated caustic). This should provide crucial information about the population of point masses in the lensing cluster.

An interesting question is whether caustic-crossing stars could be detected in galaxy-galaxy lenses. These lens systems typically have lower masses and consequently larger values of $d$ (typically ${\rm arcsec}^{-1}$), and tend to have larger values of surface mass density $\kappa_\star$ ($\simeq {\rm few} \times 0.1$) in microlenses in the vicinity of images. If we choose $\kappa_\star \simeq 0.3$, the typical peak magnification as given by \refeq{microlen_mumax} is lower by a factor of ten from the value in the cluster case; however, the source-plane width as given by \refeq{sw} is $s_w \simeq 1.5 \, {\rm kpc}$, while the number of microcaustics as given by \refeq{Nc} is $N_c \simeq 3 \times 10^5$. Hence, for the same source-plane density of stars, this makes it more probable to have a star within the caustic network. A complication in this case is the higher background level for point-source detection, which could be an observational obstacle to detecting caustic crossings in such systems.

Another interesting question is whether pulsars could be detected during caustic crossings. Let us first consider a cluster lens without any microlenses. The peak magnification at radio frequencies is limited by diffraction (see \refeq{mumaxdiff}) instead of the finite source size; the resulting magnification factor is $\mu_{\rm max, diff} \simeq 3 \times 10^7$ for the lens parameters in this paper. This magnification factor, combined with the relatively low horizon distance ($\sim 50\,{\rm kpc}$) up to which unlensed radio pulsars have been found \citep{2005AJ....129.1993M}, makes the detection of caustic-crossing radio pulsars at cosmological distances implausible. The case of X-ray pulsars is more promising, since they have been detected out to $\sim 15\,{\rm Mpc}$ \citep{2017Sci...355..817I} without the help of lensing magnification, and the peak magnification is larger (the source-size-limited value as given by \refeq{mumax-smooth} equals $\mu_{\rm t, max} \simeq 3 \times 10^9$ if the emission region is $10\,{\rm km}$ in size). Caustic-crossing X-ray pulsars behind cluster lenses should be detectable in the absence of microlensing; even in the presence of microlenses making up $\simeq 1\%$ of the surface mass density, such pulsars should be detectable intermittently during microcaustic crossings.

The study of microlensing near cluster caustics is a potentially powerful tool to study the mass function of intracluster stars or that of any possible compact objects that may be part of the dark matter. For example, these observations can probe the abundance of stellar remnants, low-mass main-sequence stars, and brown dwarfs in the intracluster stellar population. Furthermore, various subtle effects on the existing networks of stellar microcaustics can probe the presence of low-mass MACHOs as a small fraction of the dark matter over a broad mass range in a very competitive way. Applications also extend to limiting other possibilities for the constituents of dark matter that make different predictions for inhomogeneities in the surface density, such as surface mass fluctuations due to ultralight axion waves propagating through a cluster halo. Future prospects for these observations are very promising: space-based observations using HST and JWST can see very faint point sources, and high-cadence monitoring of a large number of lensing clusters using LSST can discover and characterize the brightest caustic-crossing events.

Our work was primarily motivated by the initial observations presented in \cite{2016ATel.9097....1K}. As we were finishing this paper, other manuscripts appeared in the literature that presented and interpreted subsequent observations of the same system \citep{2017arXiv170610279K, 2017arXiv170610281D}. The observations presented in \cite{2017arXiv170610279K} resolve the two image clouds on either side of the macrocritical curve, while the numerical results and scaling relations we derived in Section \ref{sec:numerical} focus on the spatially unresolved light curves of caustic-crossing events. Apart from this difference, our results on the total light curves, as well as the network of microcritical curves and caustics and the image distribution in Section \ref{sec:critanalytic}, are consistent with the simulations presented in \cite{2017arXiv170610281D}.

%%%%%%%%%%%%%%%%%%%%%%%%%%%
\acknowledgments

 We thank Timothy Brandt, Tom Broadhurst, Nick Kaiser, and Juna
Kollmeier for several helpful discussions. We also thank the referee, Prasenjit Saha, for several insightful comments and suggestions. J.M. would also like to thank Roger
Blandford and Joachim Wambsganss for discussions many years ago on the
nature of the microlensing effect near a cluster macrocaustic.

T.V. acknowledges support from the Schmidt Fellowship and the Fund for Memberships in Natural Sciences at the Institute for Advanced Study. L.D. is supported at the Institute for Advanced Study by NASA through Einstein Postdoctoral Fellowship grant number PF5-160135 awarded by the Chandra X-ray Center, which is operated by the Smithsonian Astrophysical Observatory for NASA under contract NAS8-03060. L.D. was also supported by the Natural Science Foundation of Zhejiang Province of China under LY17A050001. 
J.M. thanks the Institute for Advanced Study for their hospitality during
scientific visits; he is supported in part for this work by Spanish grant
AYA2015-71091-P.
This work utilizes gravitational lensing models produced by PIs Brada\v{c}, Natarajan \& Kneib (CATS), Merten \& Zitrin, Sharon, and Williams, and the GLAFIC and Diego groups. This lens modeling was partially funded by the HST Frontier Fields program conducted by STScI. STScI is operated by the Association of Universities for Research in Astronomy, Inc. under NASA contract NAS 5-26555. The lens models were obtained from the Mikulski Archive for Space Telescopes (MAST).
%%%%%%%%%%%%%%%%%%%%%%%%%%%

\appendix

%%%%%%%%%%%%%%%%%%%%%%%%%%%
\section{Mean magnification through a caustic crossing with microlenses}
\label{app:analytics}
%%%%%%%%%%%%%%%%%%%%%%%%%%%

Typically, in gravitational lensing theory, exact results are hard to derive except 
for the simplest lens models. In scenarios with a large number of strongly coupled 
point-mass lenses, such as the subject of this paper, it is even harder to state
exact results, and hence we resort to numerical simulations as in our Section 
\ref{sec:numerical}. In this section, we derive an exact (or nearly exact) 
analytical result for the mean magnification (over the 
realizations of the point masses) that we considered in \refsec{microlens},
as a function of source position.

The net magnification of a point source (located at $\bt y$ on the source plane) 
summed over all its images is the integral
\begin{align}
  \mu(\bt y) & = \int d^2{\bt x} \,\, \delta\left(\bt x - \bt y - \bm{\alpha}_{\rm B}(\bt x) - \bm{\alpha}_{\rm ml}(\bt x) \right) 
  = \frac{1}{(2\pi)^2} \int d^2{\bt x} \, \int d^2{\bt l}\, \, e^{- i {\bt l} \cdot \left[ \bt x - \bt y - \bm{\alpha}_{\rm B}(\bt x) \right]} e^{ i {\bt l} \cdot \bm{\alpha}_{\rm ml}(\bt x) } \mbox{.} \label{eq:mudefpt}
\end{align}
where, as in the main text, $\bm\alpha_{\rm B}$ is the deflection of the background 
lens model and $\bm\alpha_{\rm ml}$ is the deflection due to the point masses. The
latter averages to zero since the background model includes the surface mass density
in the microlenses. The mean value of the magnification is
\begin{align}
\label{eq:mumeandefl}
  \langle \mu(\bt y) \rangle & = \frac{1}{(2\pi)^2} \int d^2{\bt x} \, \int d^2{\bt l} \,\, e^{- i {\bt l} \cdot \left[ \bt x - \bt y - \bm{\alpha}_{\rm B}(\bt x) \right]} \langle e^{ i {\bt l} \cdot \bm{\alpha}_{\rm ml} } \rangle \mbox{.} 
\end{align}
Since the mean is zero, we evaluate the fluctuating part at the center of the field. 
The expectation value on the RHS is the characteristic function, $Q(\bt l)$, for the 
microlens's deflection. Using the relation between the characteristic function and the 
two-dimensional PDF $p(\bm{\alpha}_{\rm ml})$, we rewrite 
\refeq{mumeandefl} as
\begin{align}
  \langle \mu(\bt y) \rangle & = \int d^2{\bt x}\, \, p\left( \bm{\alpha}_{\rm ml} = \bt x - \bt y - \bm{\alpha}_{\rm B}(\bt x) \right) = \int d^2{\bt x}\, \, p\left( \bm{\alpha}_{\rm ml} = {\bt y}_{\rm B}(\bt x) - \bt y \right) \mbox{,} \label{eq:meanmudef}
\end{align}
i.e., the mean magnification is an integral over the image plane, with every point $\bt x$
weighted by the probability for the point-mass contribution to make up the extra angle 
needed to make $\bt x$ solve the lens equation. Transforming from the image to the 
source plane using the background lens map gives
\begin{align}
\label{eq:meanmugen}
  \langle \mu  (\bt y) \rangle & = \int d^2\bm{\alpha}_{\rm ml}\, \,p(\bm{\alpha}_{\rm ml})\, \mu_{\rm B}(\bt y + \bm{\alpha}_{\rm ml}),
\end{align}
where $\mu_{\rm B}$ is the background model's magnification.

In the simple case with widely separated macro-images, around which the background 
lens model does not vary dramatically (on scales relevant to the point-mass deflection's 
PDF $p(\bm\alpha_{\rm ml})$), the integral picks up separate contributions around each 
macro-image---the normalization of $p(\bm\alpha_{\rm ml})$ implies that the mean 
magnification is unaffected by microlensing. 

\begin{figure}[t]
\begin{center}
  \includegraphics[width=12cm]{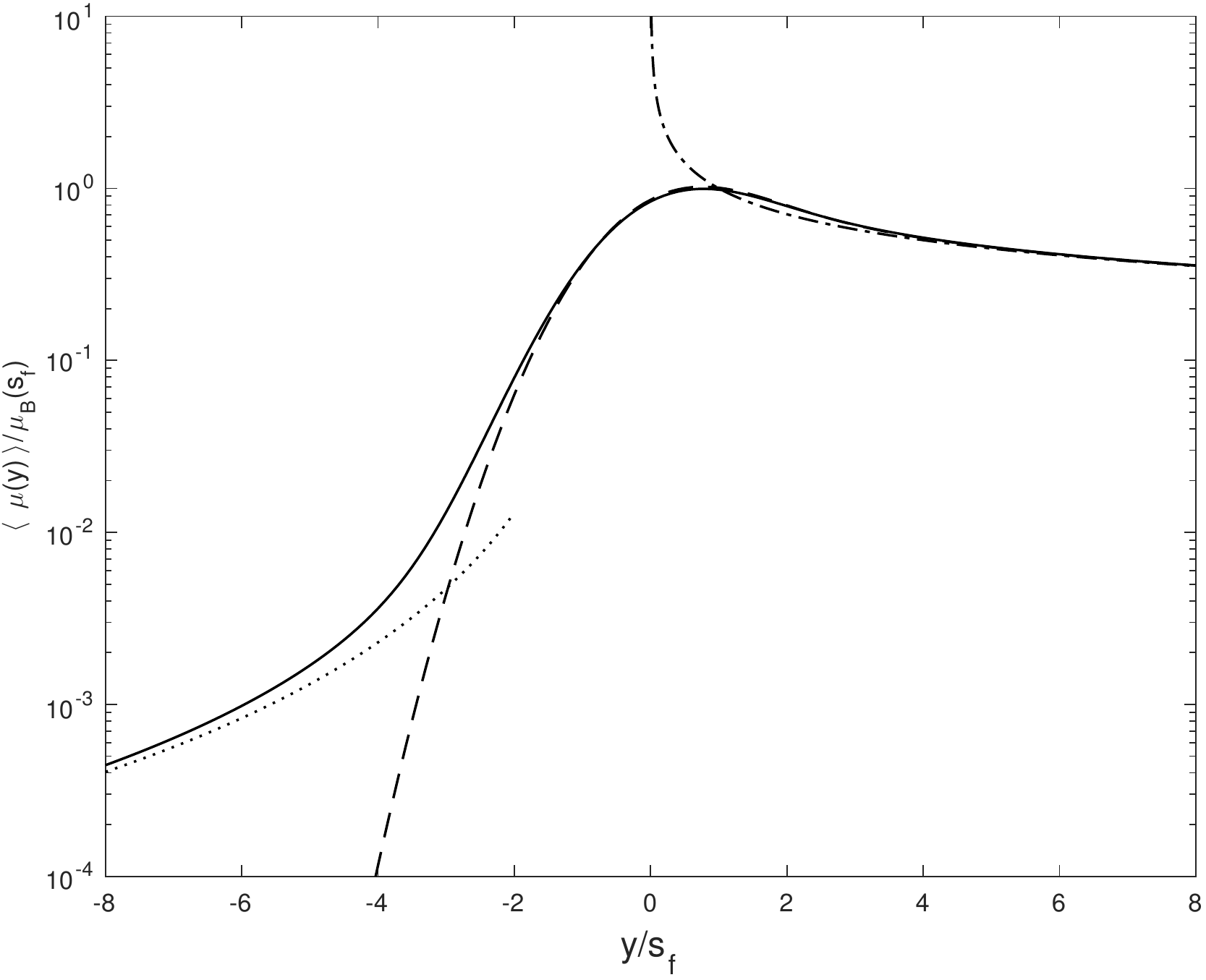}
\caption{\label{fig:meanmu} Solid line shows the mean magnification as a function 
of source position, with a population of $N_{\rm ml}=10^{6}$ microlenses in the vicinity of the 
fold model of \reffig{fold}. This is calculated using \refeq{meanmugen},
with the PDF of the point-mass deflection $p(\alpha_{\rm ml, 1})$ derived from the 
results in \cite{1986ApJ...306....2K}. The position is scaled relative to the deflection scale
$s_f$ of \refeq{sf}. The dashed-dotted curve is the background magnification, 
and the dashed and dotted curves are the limiting estimates of Equations \eqref{eq:meanmulowdef} and \eqref{eq:meanmuhighdef}, respectively.}
\end{center}
\end{figure}

A significant correction occurs when the source is near the macrocaustic, where the 
background magnification changes rapidly with source displacement. Let us define 
coordinates in the vicinity of the macrocaustic such that the mean magnification is 
given by \refeq{mu1}, which we write as $\mu_{\rm B} = A/\sqrt{y_1}$. 
Due to the one-dimensional nature of this magnification, the deflection along the 
$y_2$ direction integrates out in \refeq{meanmugen} and the relevant 
PDF is $p(\alpha_{ml, 1}) = \int d\alpha_{ml, 2} \, p(\bm{\alpha}_{\rm ml})$:
\begin{align}
  \langle \mu(\bt y) \rangle & = A \int\limits_{0}^{\infty} d\xi \,\, \frac{p\left(\alpha_{ml, 1} = \xi - y_1\right)}{\sqrt{\xi}} \mbox{.} \label{eq:meanmupl}
\end{align}
\reffig{meanmu} shows the mean magnification as a function of source 
displacement (the latter is in units of the scale $s_f$ from \refeq{sf}), 
obtained by using the PDF for the deflection $\bm{\alpha}_{\rm ml}$ from 
\cite{1986ApJ...306....2K}. The scale $s_f$ depends logarithmically on the number 
of masses $N_{\rm ml}$, the figure is for $N_{\rm ml}=10^6$.
%%%%%%%%%%%%%%%%%%%%%%%%%%%%%%%%%%%%%%%%%%%
%%%%%%%%%%%%%%%%%%%%%%%%%%%%%%%%%%%%%%%%%%%

We can analytically estimate the mean magnification in \refeq{meanmupl} in 
two limits by using the limiting behavior of the PDF $p(\bm{\alpha}_{\rm ml})$ at small and 
large deflections. The first limit is when the source position $y_1 \gtrsim -s_f$, when most 
of the integral's weight in \refeq{meanmupl} comes from the core of the 
PDF $p(\bm{\alpha}_{\rm ml})$, where the latter can be approximated by a Gaussian with 
standard deviation $s_f$. Performing the integral, we get
\begin{align}
  \frac{\langle \mu(\bt y) \rangle}{\mu_{\rm B}(s_f)} \biggr\vert_{y_1 \gtrsim -s_f} & \approx \int\limits_{0}^{\infty} \frac{d\xi}{\sqrt{2 \pi \xi}} \, \exp{\left[ - \frac{(\xi - \tilde{y}_1)^2}{2} \right]}, \, {\rm where} \,\,\,\, \tilde{y}_1 = \frac{y_1}{s_f} \\
  & = \begin{cases}
          \frac{1}{2\sqrt{\pi}}\, e^{-\tilde{y}_1^2/4}\, \sqrt{-\tilde{y}_1}\, K_{1/4}\left( \tilde{y}_1^2/4 \right), & \tilde{y}_1 \leq 0 \\
          \frac{\sqrt{\pi}}{2\sqrt{2}}\, e^{-\tilde{y}_1^2/4}\, \sqrt{\tilde{y}_1}\, \left[ I_{1/4}\left( \tilde{y}_1^2/4 \right) + I_{-1/4}\left( \tilde{y}_1^2/4 \right) \right], & \tilde{y}_1 > 0
        \end{cases}. \label{eq:meanmulowdef}
\end{align}
Here, $K$ and $I$ are modified Bessel functions.

The other regime is when the source position $y_1 \ll -s_f$, when, in order for a micro-image to exist, 
we need a large deflection from the tail of the distribution $p(\bm{\alpha}_{\rm ml})$. In this limit, 
the latter behaves like a power law. The PDF and the mean magnification are
\begin{align}
  p(\alpha_{\rm ml, 1}) & \approx \int d\alpha_{\rm ml, 2}  \frac{\theta_\star^2 \kappa_\star}{\pi} \frac{1}{(\alpha_{\rm ml, 1}^2 + \alpha_{\rm ml, 2}^2)^2} = \frac{\theta_\star^2\, \kappa_\star}{2} \frac{1}{\alpha_{\rm ml, 1}^3}, \\
  \frac{\langle \mu(\bt y) \rangle}{\mu_{\rm B}(s_f)} \biggr\vert_{y_1 \ll -s_f} & \approx \frac{\theta_\star^2\, \kappa_\star}{2\, s_f^2} \int\limits_{0}^{\infty} d\xi \, \frac{1}{\sqrt{\xi}} \frac{1}{(\xi - \tilde{y}_1)^3} = \frac{3\,\pi}{16 \, \ln{\left[ 3.05 N_{\rm ml}^{1/2} \right]} } \frac{1}{\tilde{y}_1^{5/2}}. \label{eq:meanmuhighdef}
\end{align}
\reffig{meanmu} shows the mean magnifications of Equations \eqref{eq:meanmulowdef} and 
\eqref{eq:meanmuhighdef} using dashed and dotted lines, respectively.

%%%%%%%%%%%%%%%%%%%%%%%%%%%
\section{Candidate event in MACS\,J1149+2223}
\label{app:application}
%%%%%%%%%%%%%%%%%%%%%%%%%%%

Recently, an intriguing transient (J2000 coordinates R.A. = 11:49:35.66 and decl. = 22:23:48.0) has emerged in the {\em HST}'s view of the galaxy cluster MACS\,J1149+2223 ($z_L=0.544$). As reported in \cite{2016ATel.9097....1K}, starting from 2016 April 29, the point-like transient brightened from a {\em J}-band (F125W) magnitude = $26.5 \pm 0.1$ ABmag to $25.8$, and an {\em R}-band (F606W) magnitude of $26.8 \pm 0.1$ over one month. Wide-band photometry suggests a redshifted spectrum similar to that of a B-type star. It has been interpreted as a caustic-crossing event because the transient appears to coincide with the galaxy cluster's critical curve, assuming that the source resides in the host galaxy of SN Refsdal at $z_S=1.49$~\citep{Kelly:2014mwa}.

The Frontier Fields Lens Models project\footnote{\url{https://archive.stsci.edu/prepds/frontier/lensmodels/}} provides a compilation of reconstructed lens models for MACS\,J1149+2223. These model the surface mass distribution in the galaxy cluster and produce convergence and shear maps with a resolution on the order of $\mathcal{O}(10 - 100)\,{\rm mas}$.

In \reftab{lens-model}, we apply a number of smooth lens models to the observed transient. At the coordinates of the transient, we measure the local convergence $\kappa$ and shear $\gamma$, which are expected to satisfy $\kappa+\gamma \approx 1$ for a caustic-crossing event.  In three models, {\tt Bradac}, {\tt CATS} and {\tt Merten}, the transient does not appear to be very close to the predicted critical curve, the reason for which might be low resolution or reconstruction uncertainties. In four other models, {\tt GLAFIC}, {\tt Sharon}, {\tt Williams} and {\tt Zitrin-ltm}, good coincidences (within 0.3 arcsec) are found between the transient and the cluster's critical curve. This is shown in \reffig{magplot} for three of the highest-resolution models, namely {\tt GLAFIC}, {\tt Sharon}, and {\tt Zitrin-ltm}. Those models predict a local convergence in the vicinity of the critical curve in the range $\kappa_0 = $0.77 -- 0.83. We furthermore measure the gradient of convergence and shear in these models, and find a range of values for the gradient $|\bfd|=$ 2.4 -- 5.0 ${\rm arcmin}^{-1}$ and for the combination $|d\sin\alpha| =$ 2.4 -- 4.7 ${\rm arcmin}^{-1}$. Aware of model-to-model variation, throughout this paper we adopt fiducial values for macrolensing parameters $\kappa_0=0.83$, $|\bfd|=5\,{\rm arcmin}^{-1}$ and $|d\,\sin\alpha|=4\,{\rm arcmin}^{-1}$. Despite the uncertainty in these parameters, it is verified that a smooth mass distribution on the cluster scale typically has $|\bfd|$ and $|d\,\sin\alpha|$ on the order of ${\rm arcmin}^{-1}$.
 
%%%%%%%%%%%%%%%%%%%%
\begin{table}[t]
\begin{center}
\setlength\tabcolsep{9pt}
\begin{tabular}{l|c|c|c|c|c|l}
\specialrule{.1em}{.05em}{.05em} 
 model & pixel size & $\kappa$ & $\gamma$ & $|\bfd|$ & $\left|d\,\sin\alpha\right|$ & references \\
\hline
\hline
{\tt Bradac} & $0.044''$ & 0.989 & 0.749 & -- & -- & \cite{2005AnA...437...39B,2009ApJ...706.1201B}\\
{\tt CATS} & $0.20''$ & 0.906 & 0.184 & -- & -- & \cite{2009MNRAS.395.1319J,2014MNRAS.444..268R}\\
{\tt GLAFIC} & $0.030''$ & 0.832 & 0.144 & 5.0 & 4.7 & \cite{Kawamata:2015haa,2010PASJ...62.1017O} \\
{\tt Merten} & $7.1''$ & 0.624 & 0.219 & -- & --  &\cite{2009AnA...500..681M}\\
{\tt Sharon} &  $ 0.060''$ & 0.826 & 0.152 & 4.3 & 4.0 & \cite{2007NJPh....9..447J,2014ApJ...797...48J} \\
{\tt Williams} & $0.28''$ & 0.816 & 0.182 & $\sim 9$ & $\sim 2$ & \cite{2006MNRAS.367.1209L}\\
{\tt Zitrin-ltm} & $0.065''$ & 0.774 & 0.216 & 2.4 & 2.4 &\cite{2009ApJ...703L.132Z}\\
\specialrule{.1em}{.05em}{.05em} 
\end{tabular}
\caption{\label{tab:lens-model}Predictions of the Frontier Fields Lens Models for MACS\,J1149.5+2223~\citep{2007ApJ...661L..33E,2009ApJ...707L.163S,2009ApJ...703L.132Z,2012Natur.489..406Z,2014ApJS..211...21E}, evaluated at the transient coordinates R.A. = 11:49:35.66 and decl. = 22:23:48.0 (J2000). We fix the lens redshift $z_L=0.544$ and the source redshift $z_S=1.49$. We estimate the derivatives $|\bfd|$ and $|d\,\sin\alpha|$ (in ${\rm arcmin}^{-1}$), except for models that either have insufficient resolution ({\tt Merten}) or do not predict a critical curve near the transient ({\tt Bradac} and {\tt CATS}).}
\end{center}
\end{table}
%%%%%%%%%%%%%%%%%%%%

The line of sight to the transient has an angular separation of $\sim 7''$, or equivalently, $\sim 45\,{\rm kpc}$ in projected distance, to the center of the brightest central galaxy (BCG). Models of intracluster stellar population suggest that at this proximity to a typical cluster, the line of sight should intersect the extended stellar halo of the BCG that forms as a result of past and ongoing tidal disruption~\citep{2005MNRAS.358..949Z,Puchwein:2010ec}. Below we present evidence in the {\em HST} observation for a sizable amount of intracluster stars the line of sight traverses in MACS\,J1149.5+2223.

We use HST images published by the Cluster Lensing And Supernova survey with Hubble (CLASH) collaboration\footnote{\url{http://www.stsci.edu/~postman/CLASH/Home.html}} to measure the surface brightness in the direction toward the transient. Wide-band photometry measurements come from three cameras on board the HST: ACS at optical frequencies, WFC3IR in the infrared, and WFC3UVIS in the ultraviolet. 

In those images taken in 2011, there is no evidence for the transient, and hence significant contamination to the photometry. Since the line of sight intersects one image of the host galaxy of SN Refsdal, we expect the majority of the flux to be the sum of the emission from the cluster at $z_L=0.544$ and that from the background galaxy at $z_S=1.49$. Images in the infrared, as taken by WFC3IR, appear to predominantly show the structure of the foreground cluster, with the background galaxy having only subdominant contributions at those wavelengths. On the other hand, only the structure of the background galaxy is discernible at the ultraviolet wavelengths. This implies a blue component from star formation regions in the background galaxy, in addition to a red component from the presumably aged stars in the foreground cluster. Thus, to estimate the stellar abundance on the lens plane, a separation between the two components in the integrated light is desired.

%%%%%%%%%%%%%%%%%%%%
\begin{figure}[t]
  \begin{center}
    \includegraphics[scale=0.7]{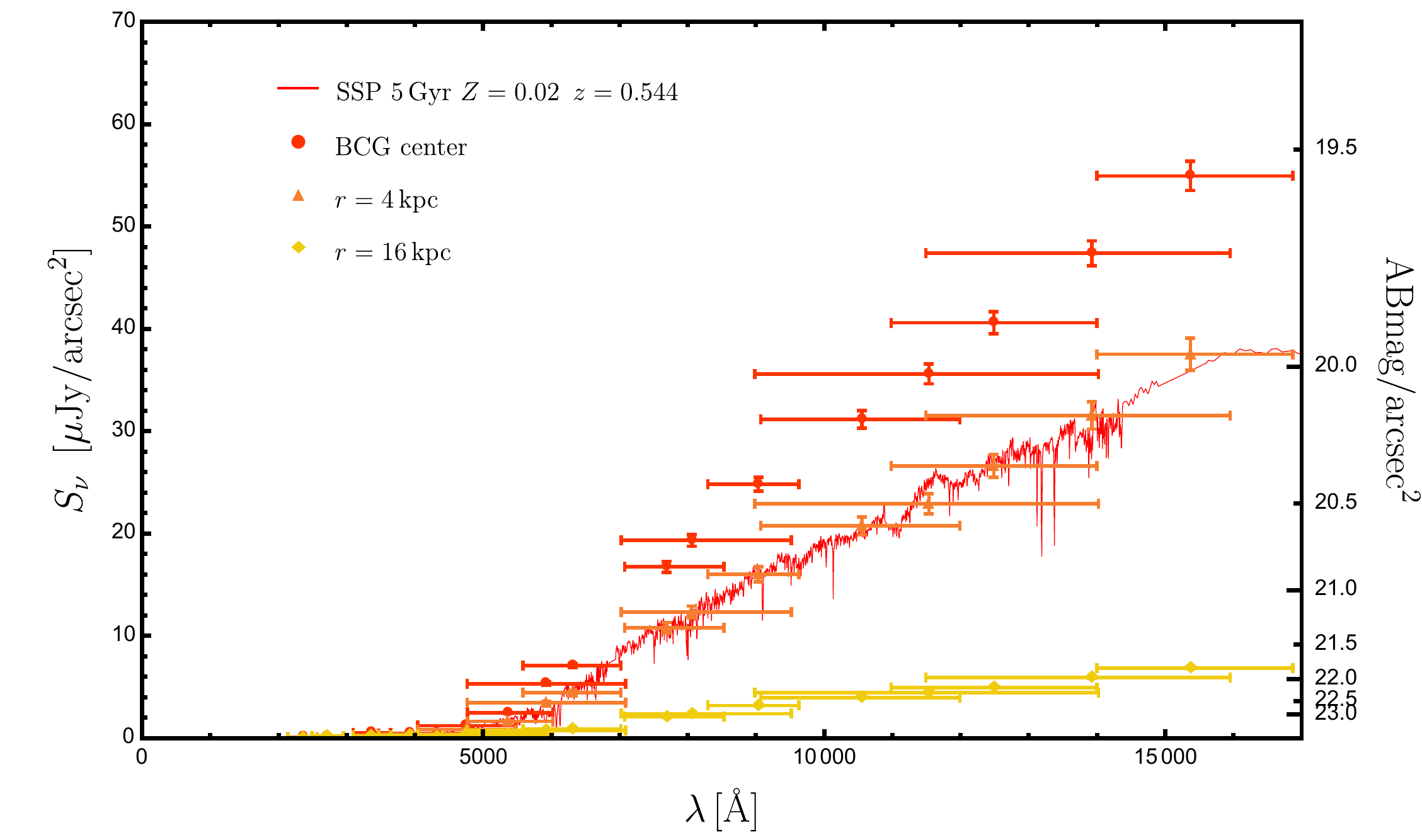}
    \includegraphics[scale=0.7]{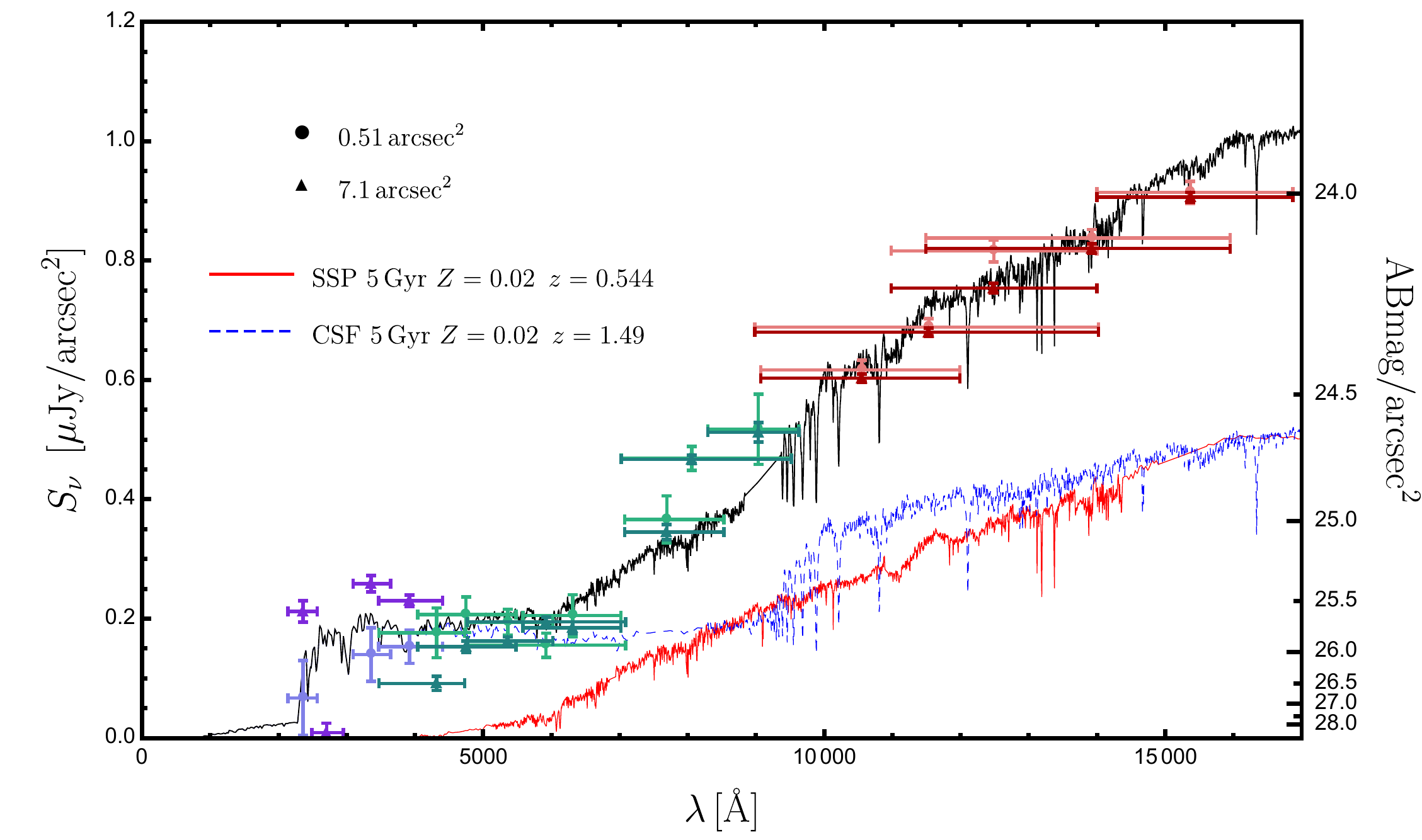}
    \caption{\label{fig:cluster_sb_HST} {\it Upper panel}: Surface brightness spectra toward the center of the BCG, a projected $4\,{\rm kpc}$ from the center, and $16\,{\rm kpc}$ from the center. The spectral shapes are well fit by an SSP $5\,{\rm Gyr}$ old with metallicity $Z=0.02$ at $z_L=0.544$. {\it Lower panel}: Surface brightness spectrum along the direction of the light of sight to the transient (RA = 11:49:35.66 and Dec = 22:23:48.0 J2000), measured from the HST images of the system MACS\,J1149.5+2223 published by CLASH collaboration. Data points correspond to the various broad-band observations by the three instruments on-board the HST, ACS (green), WFC3IR (red) and WFC3UVIS (purple), and are centered at the pivot wavelength of each band. HST images at a resolution of $65\,$milli-arcsecond are used. Vertical error bars are 1-$\sigma$ uncertainty from statistical bootstrapping; similar error bars are obtained when noises of individual pixels are assumed to be uncorrelated. Horizontal error bars indicate the FWHM of the band profile. We compare the mean surface brightness averaged over a patch of $0.51\,{\rm arcsec}^2$ (data points in lighter colors) centered at the light of sight to the transient to that over a patch of $7.1\,{\rm arcsec}^2$ (data points in darker colors). As a good fit, we over-plot the linear combination \refeq{SBspecFplusB} (solid black line) of the observed spectrum from a $5\,{\rm Gyr}$ old single stellar population (SSP) without subsequent star formation at $z_L=0.544$ (solid red line), and that from a $5\,{\rm Gyr}$ old stellar population with constant star formation (CSF) at $z_S=1.49$ (dashed blue line).}
  \end{center}
\end{figure}
%%%%%%%%%%%%%%%%%%%%

To quantify the color difference between different stellar populations, we first generated sample spectra for stellar populations of various types using the {\sc Galaxev}\footnote{\url{http://www.bruzual.org/bc03/}} code by \cite{2003MNRAS.344.1000B}, based on the \cite{2003PASP..115..763C} initial mass function (IMF) and the Padova 1994 \citep{1993A&AS..100..647B,1994A&AS..104..365F,1994A&AS..105...39F,1994A&AS..105...29F} evolutionary tracks as input. In particular, we contrast between the simple stellar populations (SSPs) that form at a single instant and have no subsequent star formation and composite stellar populations that undergo constant star formation (CSF). For CSF, we adopt a star formation rate SFR $=20\,M_\odot/{\rm yr}$, although the spectrum shape is independent of that as expected. 

Regarding our specific case, the background galaxy is at most $4.4\,{\rm Gyr}$ old and appears to host active star formation; hence, it should exhibit a spectrum of the latter type, while the stellar halo around the BCG of MACS\,J1149.5+2223 is more likely to be made up of old populations of the former type. Therefore, we may compare SSP spectra from $z_L=0.544$ to CSF spectra from $z_S=1.49$. A CSF spectrum would show a plateau above the break at around $2000\,\angstrom$, and has only a mild rise beyond $10000\,\angstrom$. The shape has only a mild dependence on age and metallicity. On the contrary, a redshifted SSP spectrum would have little emission below $5000\,\angstrom$ regardless of metallicity, unless the population is much younger than $1.4\,{\rm Gyr}$ old, which is improbable for MACS\,J1149.5+2223. Moreover, an old SSP has a much steeper rise in its spectrum beyond $5000\,\angstrom$ compared to CSF.

In the first panel of \reffig{cluster_sb_HST}, we show the surface brightness in various wide bands of the foreground cluster by aiming at regions where the foreground is undoubtedly dominant. We measure the surface brightness from the cluster's BCG and its immediate surroundings. The surface brightness is obtained by averaging over a small patch of the sky of a given area in the vicinity of the transient. We first compute the uncertainty in the averaged flux using statistical bootstrap, independently of the pixel error bars included in the HST data, which is a reasonable estimate even in the case of non-trivial noise correlation between pixels. We then directly use the photometric error bars associated with individual pixels for a comparison and obtain similar results. 

The wide-band color at a projected distances of $4\,{\rm kpc}$ and $16\,{\rm kpc}$ from the center of the cluster's BCG shows consistency with that toward the BCG core, and is well-fit by an SSP that has metallicity $Z=0.02$ and is $5\,{\rm Gyr}$ old at $z_L=0.544$. Quite differently, the spectra from typical star forming regions in the background galaxy are significantly bluer. One exception is the core of the background galaxy, whose emission is significantly redder due to a concentration of old stars; in any case, the target line of sight is far away from such a core.

Computed using the same method, the wide-band surface brightness toward the direction of the transient, for each of the broad bands in ACS, WFC3IR, and WFC3UVIS, is shown in the second panel of \reffig{cluster_sb_HST}. We also compare between a patch size of $0.51\,{\rm arcsec}^2$, which consists of 121 pixels, and a larger patch size of $7.1\,{\rm arcsec}^2$, which consists of 1681 pixels. The agreement between the two choices is consistent with a locally homogeneous diffuse emission. It is unlikely that the surface brightness back in 2011 was severely affected by the transient, which was too faint to be seen.

The color distinction between the SSP and the CSF suggests that emission from the background galaxy alone is unable to explain the measured spectrum toward the line of sight. It is not possible to fit the rise in the IR part of the spectrum with a star-forming population without violating the constraint on the UV part. Therefore, a significant fraction from the foreground cluster is necessary. In fact, the measured wide-band surface brightness $S_\nu$ can be well-fit by a linear combination of contributions from a foreground SSP of $5\,{\rm Gyr}$ old at solar metallicity, and from a background CSF at similar age and metallicity,
\ba
\label{eq:SBspecFplusB}
S_\nu & = & \Sigma_F\, f_{\nu,F} + \Sigma_B\, f_{\nu,B},
\ea
where $f_{\nu,i}$ is the spectral flux density {\it per unit stellar mass}, and $\Sigma_i$ is the stellar surface mass density along the line of sight, from the foreground $i=F$ and the background $i=B$, respectively. We found $\Sigma_F=10^9\,M_\odot/{\rm arcsec}^2$ and $\Sigma_B = 1.6\times 10^{10}\,M_\odot/{\rm arcsec}^2$. In this case, as overplotted in the second panel of \reffig{cluster_sb_HST}, the intracluster light accounts for about half of the brightness at long wavelengths. Emitting from $z_L=0.544$, it has an observed {\em I}-band (F814W) surface brightness of $26$ ABmag$/{\rm arcsec}^2$, which would translate into an observed {\em R}-band (F625W) surface brightness of $25$ ABmag$/{\rm arcsec}^2$ if the cluster were instead at $z=0.25$. When compared to the results of \cite{2005MNRAS.358..949Z} and \cite{Puchwein:2010ec}, this corresponds to the {\em R}-band surface brightness at $\sim 50\,{\rm kpc}$ away from the center of an average intracluster stellar halo at $z=0.25$, in good agreement with the projected length scale to the cluster center in our case.

The mean local convergence from intracluster stars is therefore given by
\ba
\label{eq:kappastar_fiducial}
\kappa_\star = \frac{\Sigma_F}{\Sigma_{\rm crit}} = 4\pi\,G\,\Sigma_F\,\frac{D_{LS}}{D_L\,D_S} = 0.01\,\left( \frac{\Sigma_F}{10^9\,M_\odot/{\rm arcsec}^2} \right),
\ea
which is several orders of magnitude larger than the threshold value $\kappa_c$ of \refeq{kappac}. According to \refsec{microlens}, we therefore predict that the otherwise smooth critical curve of the foreground cluster is replaced by a band of corrugated network of critical curves due to intracluster stars as microlenses.

%%%%%%%%%%%%%%%%%%%%%%%%%%%
\bibliographystyle{aasjournal}
\bibliography{reference_liang,reference_teja,reference_jordi}
%%%%%%%%%%%%%%%%%%%%%%%%%%%



\end{document}